\documentclass{scspaperproc}

\usepackage{latexsym}
\usepackage{graphicx}
\usepackage{mathptmx}

\usepackage{amsmath}
\usepackage{amsfonts}
\usepackage{amssymb}
\usepackage{amsbsy}
\usepackage{amsthm}

\usepackage[pdftex,colorlinks=true,urlcolor=blue,citecolor=black,anchorcolor=black,linkcolor=black,bookmarks=false]{hyperref}
\usepackage{accessibility}
\usepackage{booktabs}

\newif\ifblind
\blindtrue

\usepackage{hyphenat}
\newtheoremstyle{scsthe}% hnamei
{8pt}% hSpace abovei
{8pt}% hSpace belowi
{\it}% hBody fonti
{}% hIndent amounti1
{\bf}% hTheorem head fontbf
{.}% hPunctuation after theorem headi
{.5em}% hSpace after theorem headi2
{}% hTheorem head spec (can be left empty, meaning `normal')i

\theoremstyle{scsthe}

\usepackage[english]{babel}
\usepackage{placeins}   % in preamble

\IfFileExists{accessibility.sty}{\usepackage{accessibility}}{}

\makeatletter
\@ifpackageloaded{hyperref}{%
  \AtBeginDocument{\hypersetup{
    colorlinks=true,
    linkcolor=black,
    citecolor=black,
    urlcolor=blue,
    pdfborder={0 0 0}
  }}%
}{%
  \usepackage[
    colorlinks=true,
    linkcolor=black,
    citecolor=black,
    urlcolor=blue,
    pdfborder={0 0 0}
  ]{hyperref}%
}
\usepackage{tikz}
\usetikzlibrary{arrows.meta, positioning, shapes.geometric}
\makeatother
\hypersetup{
  pdftitle={ANNSIM 2026 Proceedings Paper},
  pdfauthor={Proceedings Authors},  % left generic since authorship is per-paper
  pdfsubject={ANNSIM 2026 Conference},
  pdfkeywords={}
}
\begin{document}

%***************************************************************************
% AUTHOR: AUTHOR NAMES GO HERE
% FORMAT AUTHORS NAMES Like: Author1, Author2, and Author3 (last names)
%
%		You need to change the author listing below!
%               Please list ALL authors using last name only, separate by a comma except
%               for the last author, separate with "and"
%
\SCSpagesetup{Vrinda Malhotra}

% AUTHOR: Uncomment ONE of these correct conference names.
\def\SCSconferencename{Annual Modeling and Simulation Conference}
%\def\SCSconferencename{Power Plant Simulation Conference}

% AUTHOR: Uncomment ONE of these correct conference acronyms.
\def\SCSconferenceacro{ANNSIM'26}
%\def\SCSconferenceacro{ANNSIM'25}

% AUTHOR: Set the correct year of the conference.
\def\SCSpublicationyear{2026}

% AUTHOR: Set the correct editors of the conference
\def\SCSconferenceeditors{G. Rabadi, V. Prabhu, R. Cárdenas, A. Bany Abdelnabi, J. Jabbour, and M. Germanos}

% AUTHOR: Set the correct month and dates; the dates are separated by a single minus sign
% with no spaces and no leading zeros, the month is a full name (e.g. April) with the first letter
% capitalized. For example, "April 8-13".
\def\SCSconferencedates{May 4-7}

% AUTHOR: Set the correct venue in the form "City, State, Country", for example, "Los Angeles, CA, USA".
\def\SCSconferencevenue{University of Central Florida, Orlando, Florida, USA}

% AUTHOR: Enter the title, all letters in upper case
\title{MODELING MISINFORMATION AS A COMMONS PROBLEM}

\author{Vrinda Malhotra}
\affil{George Mason University}

\maketitle

\section*{Abstract}

Misinformation often harms society not just by spreading a single false belief, but by breaking down the shared trust people rely on to evaluate what is true. This paper presents an agent-based simulation that frames trust as a collective resource and attention as a scarce private budget: when aggregate attention shifts toward low-credibility content, the trust environment degrades, making credible information harder to process and correct. Across experiments, the model produces four recurring modes: credible stability, misinformation dominance, polarization, and a mixed baseline, with distinct signatures in trust trajectories and network structure. The results separate two control problems that matter for simulation-based policy exploration: the balance of trust repair versus harm largely determines whether the system recovers or collapses, while homophily and rewiring determine whether disagreement remains integrated or separates into persistent clusters. This foundation provides a transparent testbed for comparative experiments on interventions that must address both trust restoration and structural conditions for cross-cutting exposure.

\textbf{Keywords:} agent-based simulation; attention economy; epistemic commons; adaptive networks; polarization.
%% AUTHOR:
% This is a list of no more than five keywords that will identify your paper in indices and databases (required).
% Do not use the words “computer”, “simulation”, “model”, or “modeling”, since these are all assumed.

\section{Introduction}
\label{sec:intro}

Digital information environments make attention scarce and highly leveraged: what users attend to influences what they learn, share, and reinforce, while platform-mediated networks can amplify and concentrate exposure. In such settings, misinformation not only misleads; it can degrade the epistemic environment by undermining trust, creating conditions where low-credibility narratives become easier to adopt and harder to correct~\cite{Lazer2018,Vosoughi2018}.

A large literature models belief change through interpersonal influence and bounded updating. Bounded-confidence models capture the empirically plausible idea that influence is limited when beliefs are too dissimilar, yielding consensus, fragmentation, or polarization depending on parameters and network structure~\cite{Deffuant2000,HegselmannKrause2002}. Work on bounded rationality emphasizes that attention is limited and processing is costly, shaping selective exposure and heuristic reliance~\cite{Simon1955,Kahneman2011,Sims2003}. Adaptive-network research shows that when ties co-evolve with agent states, the network can sort into segregated, metastable polarized communities~\cite{McPherson2001,GrossBlasius2008}.

Several agent-based models directly address misinformation and polarization. Del~Vicario et al.~\cite{DelVicario2016} document empirically that misinformation and credible content spread through largely separate echo chambers on social media, with homophily driving segregation. Tambuscio et al.~\cite{Tambuscio2015} model fact-checking as a competing contagion, showing that network topology and corrective density jointly determine whether false narratives are suppressed. Sasahara et al.~\cite{Sasahara2021} show that selective unfollowing accelerates echo chamber formation even among initially heterogeneous agents. These models capture important aspects of information spread and network sorting, but they treat the epistemic environment, that is, the shared pool of trusted sources and credible signals that a population draws on, as a fixed backdrop rather than a depletable resource shaped by collective behavior.

The knowledge commons literature suggests a different framing. Hess and Ostrom~\cite{HessOstrom2007} identify shared information environments as exhibiting commons-like properties: they are collectively produced, difficult to exclude individuals from, and vulnerable to degradation when overused or neglected. Frischmann et al.~\cite{Frischmann2014} extend this to knowledge infrastructure more broadly, arguing that governance failures arise precisely when individually rational use patterns degrade the shared resource. Neither Ostrom nor Hardin applies this logic to epistemic trust directly; their canonical cases are fisheries, irrigation systems, and grazing land. The extension to trust is our own analytical move, motivated by a structural analogy: like a fishery, trust in the information environment is non-excludable (everyone is affected by aggregate misinformation exposure), subject to depletion through overuse (mass consumption of low-credibility content degrades the shared signal), and not automatically self-replenishing. This paper operationalizes that analogy computationally, closing the loop between individual attention choices and the aggregate trust stock those choices produce.

This paper fills that gap. By coupling a depletable trust stock to individual attention allocation, credibility learning, and network updates, the model creates a macro--micro feedback: each agent's attention choice shifts aggregate trust, which in turn changes the cognitive cost of credible engagement for everyone. This collective-action structure can generate collapse, recovery, and polarization as distinct emergent regimes, and opens a path to studying governance levers that operate at the commons level rather than on individual agents. The model serves as a controlled simulation platform; we characterize its regime structure and provide verification and validation evidence to support its reliability as a foundation for future governance and resilience work.

\textbf{Contributions.} This paper contributes (1) a minimal agent-based simulation that couples a global trust stock, trust-conditioned attention allocation, bounded-confidence credibility learning, and adaptive-network rewiring into a closed feedback system; (2) a verification suite including differential baseline reductions and structural unit tests that check invariants and component behaviors; (3) validation via perturbation experiments (misinformation surge shock) and a stress test contrasting adaptive versus random rewiring; and (4) regime and phase-structure analyses showing that trust equilibria and polarization can be controlled by different parameter families, implying distinct intervention levers.

\begin{figure}[!t]
\centering
\begin{tikzpicture}[
  scale=0.9, transform shape,
  node distance=32mm and 32mm,
  every node/.style={font=\small},
  box/.style={draw, rounded corners, align=center, inner sep=5pt,
              minimum height=10mm, text width=32mm},
  edge/.style={->,semithick,shorten >=4pt,shorten <=4pt},
  lab/.style={font=\footnotesize, inner sep=1pt, fill=white, text=black},
  >={Latex[length=2mm]}
]
% ===== Nodes =====
\node[box] (imit) {\textbf{Preference imitation}\par{\footnotesize (update $\eta_i$)}};
\node[box, right=of imit] (prices) {\textbf{Trust-based prices}\par{\footnotesize (compute $p_g,p_m$)}};
\node[box, right=of prices] (alloc) {\textbf{Attention allocation}\par{\footnotesize (choose $I_g,I_m$)}};
\node[box, below=18mm of prices] (sig) {\textbf{Information signals}\par{\footnotesize (content exposure)}};
\node[box, right=of sig] (Tg) {\textbf{Global trust}\par{\footnotesize (update $T$)}};
\node[box, below=18mm of sig] (cred) {\textbf{Credibility learning}\par{\footnotesize (update $c_i$)}};
\node[box, right=of cred] (Ti) {\textbf{Local trust}\par{\footnotesize (compute $T_i$)}};
\node[box, below=26mm of imit] (rew) {\textbf{Adaptive rewiring}\par{\footnotesize (update ties)}};
% ===== Links =====
% Top row
\draw[edge] (imit.east) -- node[lab, pos=.55, above=2pt] {updated preferences} (prices.west);
\draw[edge] (prices.east) -- node[lab, pos=.55, above=2pt] {trust-based costs} (alloc.west);
% Allocation -> Signals
\draw[edge] (alloc.south) to[out=-150,in=50,looseness=1.15]
  node[lab, right=2pt, pos=0.55] {attention pattern} (sig.north);
% Signals -> Credibility
\draw[edge] (sig.south east) to[out=-120, in=90]
  node[lab, sloped, above, pos=.55] {exposure to content} (cred.north);
% Aggregate exposure -> Global trust
\draw[edge] (alloc.south) to[out=-90, in=90, looseness=1.0]
  node[lab, above, pos=.55] {aggregate exposure} (Tg.north);
% Global trust -> Local trust
\draw[edge] (Tg.south) -- node[lab, pos=.5, above=2pt] {global trust signal} (Ti.north);
% Local trust -> Prices (feedback)
\draw[edge] ([yshift=1pt]Ti.north east) .. controls +(0,30mm) and +(10,30mm) .. node[lab, sloped, above, pos=.4] {local trust} ([yshift=-1pt]prices.north east);
% Credibility -> Rewiring
\draw[edge] (cred.west) -- node[lab, pos=.5, above=2pt] {credibility profile} (rew.east);
% Rewiring -> Preference
\draw[edge] (rew.north) -- node[lab, pos=.5, above=2pt] {social neighborhood} (imit.south);
% Rewiring -> Credibility (network feedback; keep short + curved to avoid crossings)
\draw[edge] (rew.north east) to[out=25,in=205,looseness=1.1]
  node[lab, above, pos=.55] {updated network} (cred.south west);
\end{tikzpicture}
\caption{Conceptual information flow in the updated misinformation--commons model. Preferences and trust-based prices determine individual attention splits. Attention drives exposure, which updates both global trust and agent credibility. Local trust and adaptive rewiring feed back by reshaping neighborhoods and network structure.}
\label{fig:model_overview}
\end{figure}
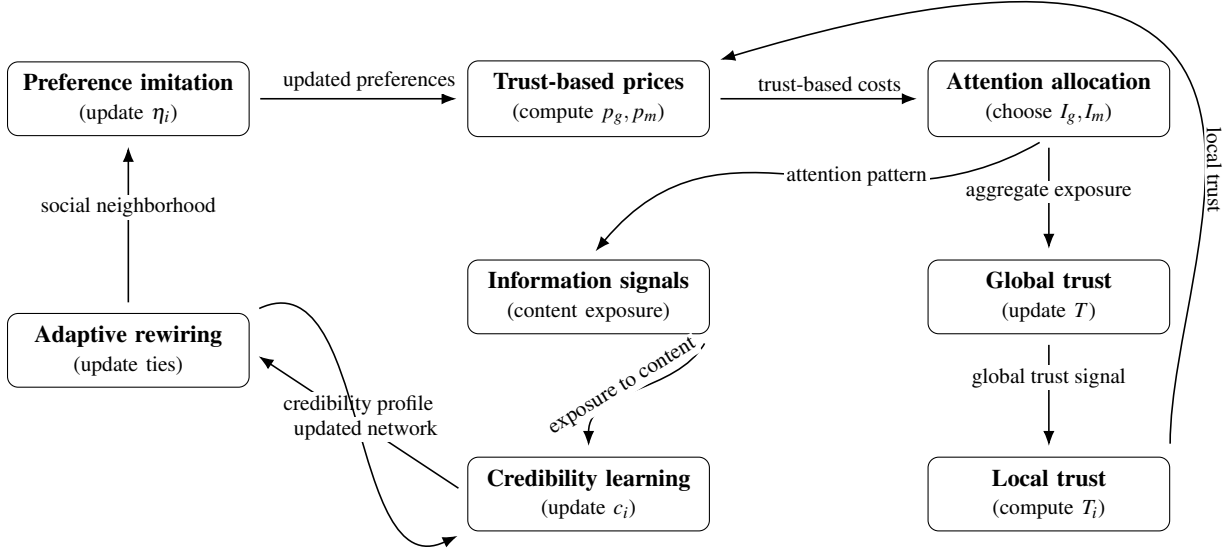

% ============================================================

\normalsize
\section{Methods}
\label{sec:methods}

The design follows the generative social science approach of explaining macroscopic outcomes via transparent micro-mechanisms~\cite{Epstein2013}. Individual building blocks are drawn from established literature: bounded-confidence opinion dynamics~\cite{Deffuant2000,HegselmannKrause2002}, social influence models~\cite{DeGroot1974,FriedkinJohnsen2011}, and adaptive networks with homophily~\cite{GrossBlasius2008,HolmeNewman2006,McPherson2001}; and assembled into a closed feedback system (Fig.~\ref{fig:model_overview}). The novel element is the commons coupling: attention is private, but its aggregate pattern depletes or repairs a shared trust stock, creating an externality that feeds back through trust-conditioned cognitive prices to every agent’s next attention decision.

\subsection{Agent States and Network Substrate}
Agents are embedded in a social network that structures repeated exposure and local learning. We use a small-world substrate to capture high clustering with occasional long-range ties, a common approximation for social exposure networks~\cite{WattsStrogatz1998}. At time $t$, each agent $i$ has: (i) a credibility belief $c_i^t\in[0,1]$, (ii) an attention split between credible and misinformation streams $(g_i^t,m_i^t)$ constrained by $g_i^t+m_i^t=1$, and (iii) a preference/bias parameter $\eta_i^t\in[0,1]$. The environment maintains a global trust stock $T^t\in[0,1]$.

\subsection{Global Trust Dynamics (Repair vs.\ Harm)}
We treat trust as a population-level stock that can be repaired by credible exposure and harmed by misinformation exposure. This reflects empirical evidence that misinformation can spread widely and undermine trust in information ecosystems~\cite{Vosoughi2018,Lazer2018}. Trust evolves via a logistic balance between repair and harm:
\begin{equation}
\frac{dT}{dt} =
\alpha_{\mathrm{up}}\,\bar{I}_g(1-T)\;-\;\beta_{\mathrm{down}}\,\bar{I}_m\,T,
\label{eq:trust}
\end{equation}
where $\alpha_{\mathrm{up}}$ is the repair strength, $\beta_{\mathrm{down}}$ is the harm strength, $N$ is the network size, and $\bar{I}_g,\bar{I}_m$ are mean credible and misinformation attention, respectively:
\begin{equation}
\bar{I}_g^{\,t}=\frac{1}{N}\sum_i g_i^t, \qquad \bar{I}_m^{\,t}=\frac{1}{N}\sum_i m_i^t.
\label{eq:mean_attention}
\end{equation}
The $(1-T)$ term yields diminishing returns to repair near $T\approx 1$, while the $T$ term makes harm most potent when trust still exists, capturing asymmetric fragility of trust.

\subsection{Local Trust Perception and Heterogeneity}
Trust is experienced locally. Neighborhood composition and homophily can generate spatially uneven trust environments even when the global trust stock is shared. We represent local heterogeneity by a convex combination of the global trust stock and a neighborhood signal:
\begin{equation}
T_i = (1-\xi)\,T + \xi\,T_{\mathrm{nbr},i},
\label{eq:localtrust}
\end{equation}
where $T_i$ is the trust stock of agent $i$, $T_{\mathrm{nbr},i}$ is the average credibility among agent $i$’s neighbors or $Nbr(i)$ and $\xi\in[0,1]$ controls the strength of heterogeneity:
\begin{equation}
T_{\mathrm{nbr},i}=\frac{1}{|Nbr(i)|}\sum_{j\in Nbr(i)} c_j.
\label{eq:neighbor_signal}
\end{equation}
This mirrors social influence perspectives where global context and local interpersonal signals jointly shape perceptions~\cite{DeGroot1974,FriedkinJohnsen2011}.

\subsection{Attention Allocation Under Trust-Dependent Cognitive Costs}
Attention is scarce, and information processing is costly. We operationalize this by introducing trust-dependent cognitive ``prices'' for processing credible and misinformation streams:
\begin{equation}
p_g(T)=\frac{\lambda}{T+\epsilon}, \qquad
p_m(T)=\frac{\lambda}{1-T+\epsilon},
\label{eq:prices}
\end{equation}
where $\lambda$ scales price sensitivity and $\epsilon>0$ prevents singularities. This implements a bounded-rationality interpretation in which trust conditions change the effective cognitive cost of processing information~\cite{Simon1955,Kahneman2011,Sims2003}. Under higher trust, credible information becomes easier to process (lower $p_g$) while misinformation becomes comparatively more costly (higher $p_m$), and vice versa under low trust.

Agents allocate attention by minimizing cognitive cost subject to a unit attention budget:
\begin{equation}
\min_{g_i,m_i}\; p_g(T_i)\,g_i + p_m(T_i)\,m_i
\quad \text{s.t.}\quad g_i+m_i=1.
\label{eq:attn_opt}
\end{equation}
We then incorporate a preference parameter $\eta_i$ to represent a bias toward misinformation versus credible attention, yielding the final allocations:
\begin{equation}
g_i = (1-\eta_i)\frac{p_m(T_i)}{p_g(T_i)+p_m(T_i)},\qquad
m_i = \eta_i\frac{p_g(T_i)}{p_g(T_i)+p_m(T_i)}.
\label{eq:alloc}
\end{equation}
$\eta_i$ can alternatively be interpreted as propensity to engage in low-credibility content; high $\eta_i$ shifts mass toward $m_i$ holding prices fixed.

\subsection{Credibility Updating via Bounded Confidence}
Credibility updates occur only during interactions. We use a bounded-confidence rule: agent $i$ interacts with neighbor $j$ only if their beliefs are sufficiently close:
\begin{equation}
|c_j-c_i|<\epsilon_{\mathrm{conf}}.
\label{eq:bc}
\end{equation}
If interaction occurs, both shift credibility toward each other:
\begin{equation}
c_i \leftarrow c_i + \mu(c_j-c_i), \qquad
c_j \leftarrow c_j + \mu(c_i-c_j),
\label{eq:cred_update}
\end{equation}
where $\mu\in(0,1)$ controls adjustment strength. This follows canonical bounded-confidence formulations that generate consensus, fragmentation, or polarization depending on parameters and network structure~\cite{Deffuant2000,HegselmannKrause2002}.

\subsection{Preference Dynamics: Imitation and Mutation}
Preferences evolve through social imitation with innovation noise:
\begin{equation}
\eta_i^{t+1} = (1-\rho_\eta)\eta_i^{t} + \rho_\eta\,\bar{\eta}_{N(i)}^{\,t} + \mathcal{N}(0,\sigma_\eta),
\label{eq:eta}
\end{equation}
where $\rho_\eta$ is the imitation weight, $\sigma_\eta$ controls mutation/innovation, $\bar{\eta}_{N(i)}^{\,t}$ denotes the mean preference of $i$’s neighbors at time $t$, where $\rho_\eta$ is the imitation weight, $\sigma_\eta$ controls mutation/innovation, and
$\mathcal{N}(0,\sigma_\eta)$ denotes an additive zero-mean Gaussian noise term with standard deviation $\sigma_\eta$
(applied independently to each agent at each time step), capturing exogenous preference shocks/innovation. This resembles linear social influence models in which individuals partially anchor on their prior state while moving toward local averages~\cite{DeGroot1974,FriedkinJohnsen2011}. Imitation strength is a key driver of regime stability and nonlinear transitions, consistent with sensitivity results.

\subsection{Adaptive Network Dynamics: Homophily and Rewiring}
People tend to form and maintain connections with others who share similar views, a well-documented tendency called \emph{homophily}~\cite{McPherson2001}.
In the model, this is operationalized as \emph{adaptive rewiring}: at each step, an agent may drop a connection to a dissimilar neighbor (measured by preference $\eta$) and replace it with a link to a more similar agent elsewhere in the network.
The strength of this pull toward similarity is controlled by $\beta_{\mathrm{homophily}}$: when $\beta_{\mathrm{homophily}}$ is small, rewiring is nearly random; when it is large, agents strongly favor like-minded ties.

The consequence is a feedback loop between social structure and beliefs.
As agents with similar preferences cluster together, they are exposed primarily to similar credibility signals, which reinforces their existing beliefs through the Deffuant learning rule (Section~2.5) and narrows the range of preferences they imitate (Section~2.6).
Over time this produces self-sorted neighborhoods (like echo chambers) even without any external manipulation.
Prior work on adaptive networks shows that this coevolution of ties and states can produce sharp, discontinuous transitions to fragmented, polarized structures~\cite{GrossBlasius2008,HolmeNewman2006}; our model captures the same mechanism in the context of information credibility and trust.

% ============================================================
\normalsize
\section{Experimental Results}

\label{sec:results}

We summarize simulation outcomes using (i) the global trust stock $T(t)$, (ii) population mean attention $\bar{I}_g(t)$ and $\bar{I}_m(t)$,
(iii) the evolving preference distribution $\eta_i(t)$, and (iv) network-structure indicators under rewiring.
Across runs, we observe four qualitatively distinct regimes: \emph{credible}, \emph{misinformation}, \emph{polarized}, and a \emph{baseline}
comparison; that differ both in long-run trust and in whether credibility, attention, and preferences remain well-mixed or fragment into
persistent clusters. We report regime signatures using network snapshots (Panel Set A; Fig.~\ref{fig:panel_networks}) and non-network
diagnostics (Panel Set B; Fig.~\ref{fig:panel_metrics}). To support interpretability and reliability as a simulation testbed, we embed
verification and validation evidence directly in the results via baseline reductions, unit tests, perturbation experiments, stress tests,
and parameter dependence diagnostics (Figs.~\ref{fig:verification_baselines_3x2}--\ref{fig:phase_and_regime}, Table~\ref{tab:verification_tests}).

\subsection{Regimes and operating conditions}
Regimes are induced by the joint settings of repair and harm in the trust stock and by the strength of rewiring and homophily.
High repair ($\alpha_{\mathrm{up}}$) and weak harm ($\beta_{\mathrm{down}}$) support recovery toward high trust, while weak repair and strong harm
produce collapse. Under intermediate repair/harm, homophily ($\beta_{\mathrm{homophily}}$) and rewiring determine whether the system remains mixed
(baseline) or separates into polarized clusters. Table~\ref{tab:regime_params} summarizes the qualitative parameter patterns used to generate
representative trajectories and snapshots (directionality and relative strength rather than fixed numeric values).

\begin{table}[t]
  \caption{Quantitative parameter ranges used to instantiate representative regimes.
  Ranges are binned from the explored sweep domains in Fig.~7 (repair/harm, homophily) and Fig.~8 (rewiring).
  Note: by Eq.~(7), larger $\eta$ increases misinformation attention $m_i$ (misinformation-leaning).}
  \label{tab:regime_params}
  \centering
  \small
  \setlength{\tabcolsep}{4pt}
  \renewcommand{\arraystretch}{0.95}
  \begin{tabular}{@{}p{1.95cm}p{1.55cm}p{1.55cm}p{2.55cm}p{1.55cm}p{1.55cm}@{}}
    \toprule
    Regime &
    $\alpha_{\mathrm{up}}$ (repair) &
    $\beta_{\mathrm{down}}$ (harm) &
    Preference $\eta$ (initial / target pattern) &
    $\beta_{\mathrm{homophily}}$ &
    $p_{\mathrm{rewire}}$ \\
    \midrule

    Credible &
    $[2.0,\,3.0]$ (high) &
    $[0.5,\,1.5]$ (low) &
    $\eta \in [0.0,\,0.3]$ (credible-leaning) &
    $[0,\,2]$ (low) &
    $[0.0,\,0.1]$ (low) \\

    \midrule
    Misinformation &
    $[0.3,\,1.0]$ (low) &
    $[3.5,\,5.0]$ (high) &
    $\eta \in [0.7,\,1.0]$ (misinfo-leaning) &
    $[0,\,2]$ (low) &
    $[0.0,\,0.1]$ (low) \\

    \midrule
    Polarized &
    $[1.0,\,2.0]$ (moderate) &
    $[1.5,\,3.0]$ (moderate) &
    bimodal mix (e.g., mass near $\eta\approx0.2$ and $\eta\approx0.8$) &
    $[8,\,12]$ (high) &
    $[0.2,\,0.4]$ (intermediate) \\

    \midrule
    Baseline &
    $[1.0,\,2.0]$ (moderate) &
    $[1.5,\,3.0]$ (moderate) &
    $\eta \approx 0.5 \pm 0.05$ (neutral) &
    $[2,\,6]$ (moderate) &
    $[0.1,\,0.2]$ (low--mid) \\

    \bottomrule
  \end{tabular}
\end{table}

\subsection{Regime signatures in trajectories, distributions, and networks}
\begin{figure*}[!t]
  \centering
  \includegraphics[width=0.48\textwidth]{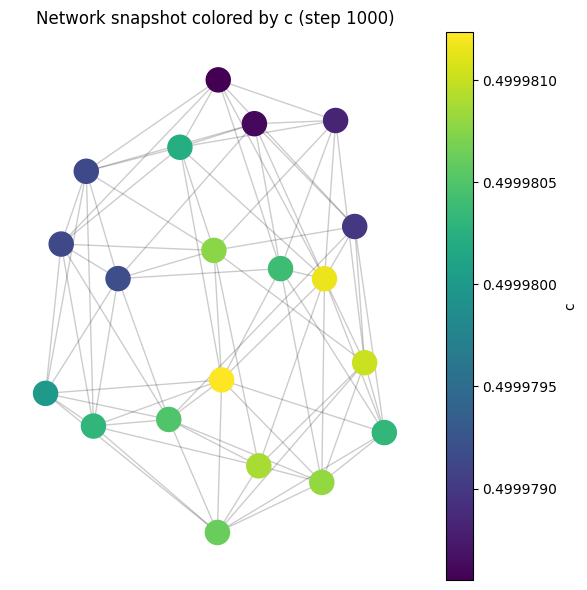}\hfill
  \includegraphics[width=0.48\textwidth]{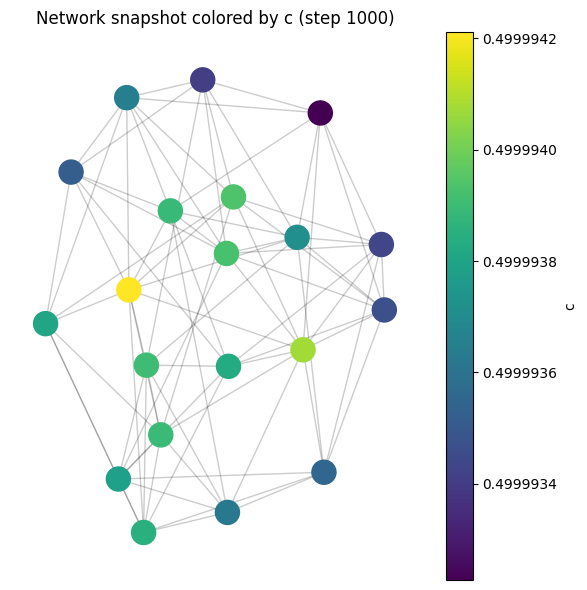}\\[3mm]
  \includegraphics[width=0.48\textwidth]{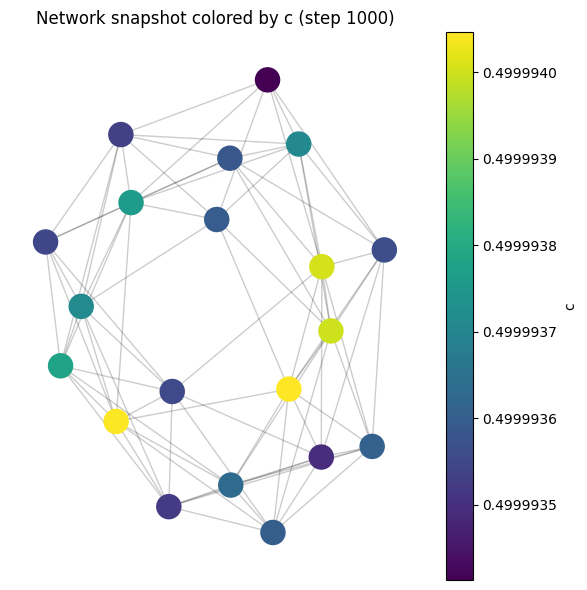}\hfill
  \includegraphics[width=0.48\textwidth]{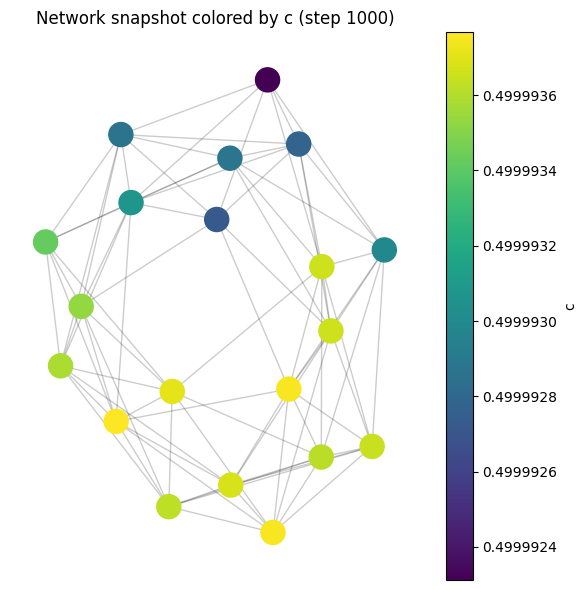}

  \caption{Network snapshots colored by credibility $c_i$: Credible (top-left), Misinformation (top-right), Polarized (bottom-left), Baseline (bottom-right). Warmer colors indicate higher credibility. Network snapshots for local trust $T_i$ and preference $\eta_i$ appear in Appendix Figs.~\ref{fig:panel_networks_full_r2}--\ref{fig:panel_networks_full_r3}.}
  \label{fig:panel_networks}
\end{figure*}

\begin{figure*}[t]
  \centering
  % Phase plots only (line plots moved to Appendix, Fig.~\ref{fig:panel_phase_full})
  \includegraphics[width=0.48\textwidth]{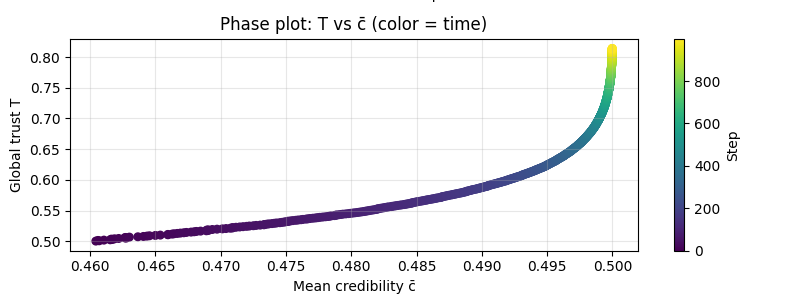}\hfill
  \includegraphics[width=0.48\textwidth]{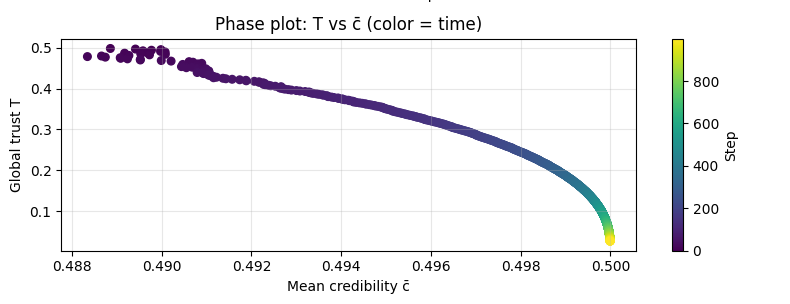}\\[3mm]
  \includegraphics[width=0.48\textwidth]{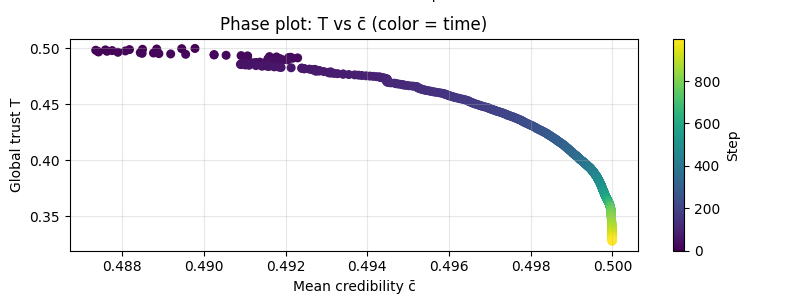}\hfill
  \includegraphics[width=0.48\textwidth]{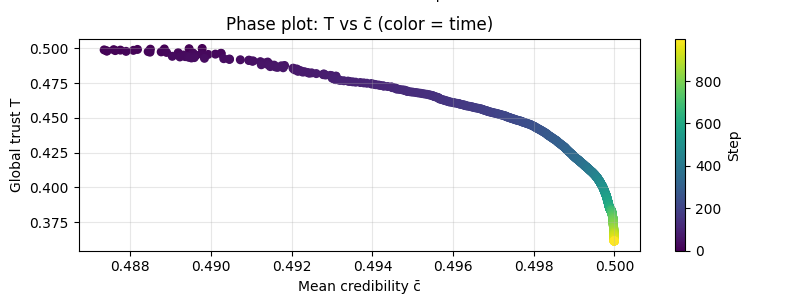}

  \caption{Phase portraits ($T$ vs.\ $\bar{c}$) for all four regimes: Credible (top-left), Misinformation (top-right), Polarized (bottom-left), Baseline (bottom-right). Each trajectory shows how the system evolves in trust--credibility space. Corresponding time-series line plots appear in Appendix Fig.~\ref{fig:panel_phase_full}; distributions and correlation diagnostics in Appendix Figs.~\ref{fig:panel_metrics_full}--\ref{fig:panel_metrics_full_r3}.}
  \label{fig:panel_metrics}
\end{figure*}

\paragraph{How to read the figures.}
Fig.~\ref{fig:panel_networks} shows network snapshots colored by credibility $c_i$ for all four regimes; Fig.~\ref{fig:panel_metrics} shows the corresponding $T$--$\bar{c}$ phase plots.
Full trajectory time series appear in Appendix Fig.~\ref{fig:panel_phase_full}; credibility and preference distributions in Appendix Fig.~\ref{fig:panel_metrics_full_r2}; additional network snapshots colored by local trust $T_i$ and preference $\eta_i$ in Appendix Figs.~\ref{fig:panel_networks_full_r2}--\ref{fig:panel_networks_full_r3}.
All panels show representative single runs; regime parameter ranges are given in Table~\ref{tab:regime_params}.

% =========================
% Panel Set A: Network Credibility Row (1x4, main body)
% Full 3x4 panel (credibility, local trust, preference) in Appendix (Fig.~\ref{fig:panel_networks_full})
% =========================

\paragraph{Trust dynamics and attention across regimes.}
The credible and misinformation regimes represent opposite poles of the trust dynamics.
In the credible regime, high repair rates and credible-leaning preferences push trust to a high stable level; rising trust then lowers the cognitive price of credible content (Eq.~\ref{eq:prices}), reinforcing the repair loop and contracting the credibility distribution toward consensus (Appendix Fig.~\ref{fig:panel_metrics_full_r2}, Appendix Fig.~\ref{fig:panel_phase_full}).
In the misinformation regime the reverse occurs: low repair and high harm rates collapse trust, which flips relative prices and concentrates attention on misinformation, locking the system into a low-trust basin.
The $T$--$\bar{c}$ phase plots (Fig.~\ref{fig:panel_metrics}) make these two attractors visible as trajectories converging to opposite regions of the phase space.
The polarized and baseline regimes occupy intermediate trust levels, but differ in dispersion: the polarized case exhibits higher variance in both trust and credibility, with bimodal distributions reflecting the coexistence of two distinct belief communities (Appendix Fig.~\ref{fig:panel_metrics_full_r2}).

\paragraph{Network structure and echo chambers.}
Network snapshots (Fig.~\ref{fig:panel_networks}) reveal distinct structural signatures across regimes.
The credible and misinformation regimes remain comparatively well-mixed: the trust--attention feedback operates uniformly across the population, so no strong spatial segregation in credibility emerges even at extreme trust levels.
The polarized regime, by contrast, shows explicit community formation: high- and low-credibility nodes cluster into separate neighborhoods, driven by homophily-based rewiring and reinforced by bounded-confidence learning once cross-cutting ties are severed (see Appendix Figs.~\ref{fig:panel_networks_full_r2}--\ref{fig:panel_networks_full_r3} for the same pattern in local trust and preferences).
The baseline lies between these extremes, moderate clustering with meaningful cross-cutting exposure, and serves as the reference trajectory for verification and validation tests below.

% =========================
% Panel Set B: Validation/sensitivity row (1x4, main body)
% Full 4x4 panel (rows 1-3) in Appendix (Fig.~\ref{fig:panel_metrics_full})
% =========================

\subsection{Verification embedded in results: baseline reductions and unit tests}
To verify internal consistency before interpreting emergent regimes, we use differential baseline reductions that remove specific feedback
channels while preserving remaining mechanisms (Fig.~\ref{fig:panel_metrics_full_r3}, Appendix~\ref{sec:appendix_panel_b}).

\begin{figure*}[!t]
  \centering
  % Row 1: correlation
  \includegraphics[height=5cm,width=0.48\textwidth]{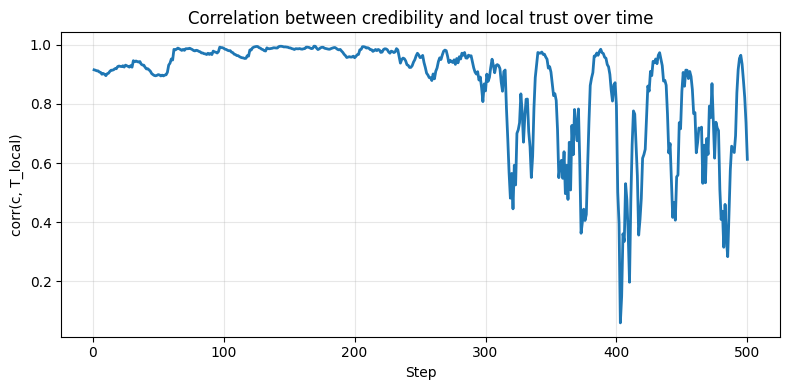}\hfill
  \includegraphics[height=5cm,width=0.48\textwidth]{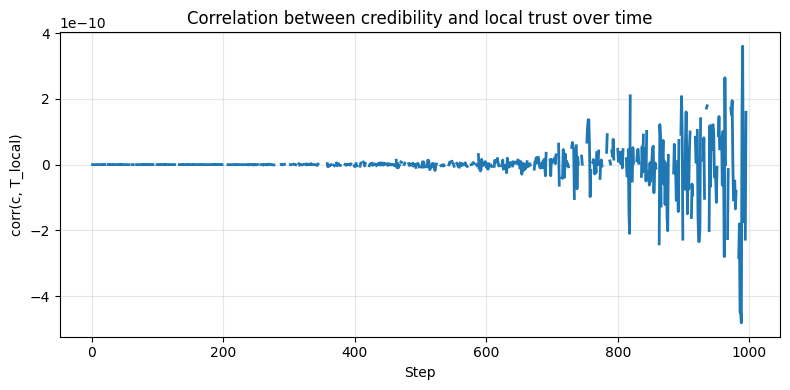}\\[3mm]
  % Row 2: phase
  \includegraphics[height=7cm,width=0.48\textwidth]{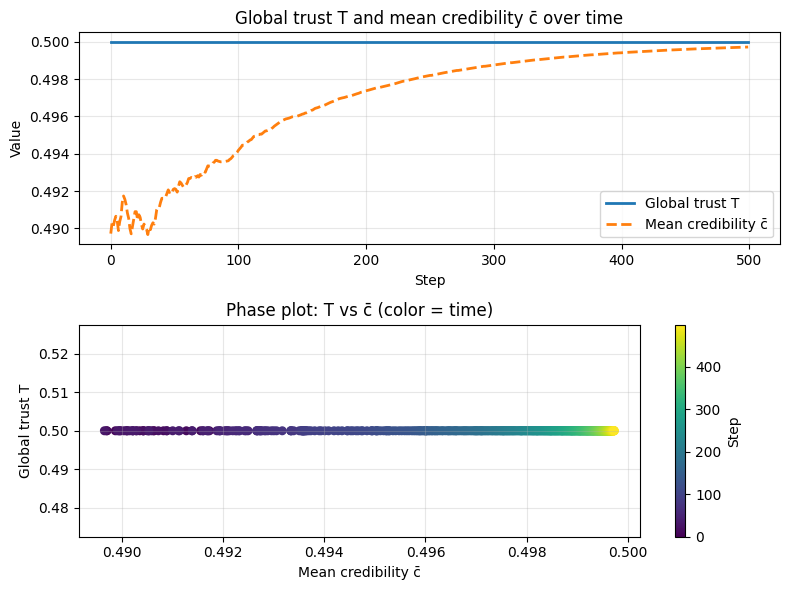}\hfill
  \includegraphics[height=7cm,width=0.48\textwidth]{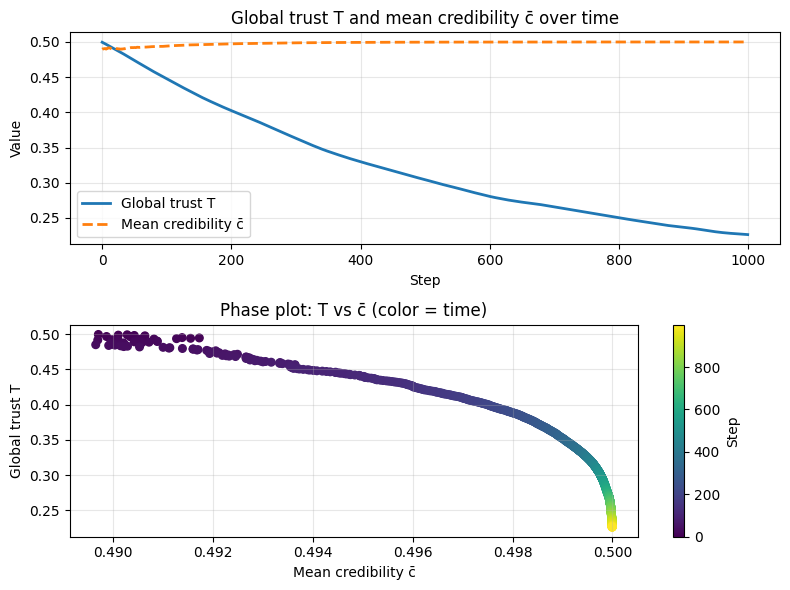}

  \caption{Verification baselines. \textbf{Left column:} fixed global trust $T(t)\equiv \bar{T}$ (commons feedback removed). \textbf{Right column:} homogeneous trust $T_i(t)\equiv T(t)$ (local heterogeneity removed). \textbf{Row 1 and 2:} correlation between credibility and local trust over time. \textbf{Row 3} phase plot of $T$ vs.\ mean credibility $\bar{c}$.}
  \label{fig:verification_baselines_3x2}
\end{figure*}

\paragraph{Baseline I: fixed global trust $T(t)\equiv \bar{T}$.}
Fixing the global trust stock and removing trust-dependence from prices isolates learning and network mechanisms from commons feedback.
As intended, Fig.~\ref{fig:verification_baselines_3x2} shows a near-zero credibility--trust correlation (left column, top) and a degenerate $T$--$\bar{c}$ phase portrait (left column, bottom),
confirming that the trust--credibility coupling has been cleanly removed.

\paragraph{Baseline II: homogeneous trust $T_i(t)\equiv T(t)$.}
Disabling local heterogeneity removes cross-sectional variation in $T_i$ while retaining global trust dynamics.
Figure~\ref{fig:verification_baselines_3x2} (right column) shows that $T(t)$ evolves but the credibility--local-trust correlation is near zero (top) because
$T_i$ has no variance, confirming that state--neighborhood coupling in the full model is produced by heterogeneity (Eq.~\ref{eq:localtrust}).

\paragraph{Structural verification via unit tests.}
We additionally implement a suite of Python tests using the library pytest, in order to assert invariants and expected component behaviors. Table~\ref{tab:verification_tests} (Appendix~\ref{sec:appendix_tests}) summarizes
coverage, including boundedness, nonnegativity, bounded-confidence gating, monotonic trust response under extreme exposures, and edge-count
preservation under rewiring. The source code is available in our~\href{https://github.com/TheChirpyWitch/misinformation_as_a_commons_problem}{project repository}~\cite{repo}.

\subsection{Validation embedded in results: shock, stress, and parameter dependence}
We validate the model by checking whether it responds plausibly to exogenous perturbations and whether structural mechanisms behave as intended
(Fig.~\ref{fig:panel_metrics}).

\begin{figure*}[!t]
  \centering
  % Row 1: Shock recovery | C-T correlation
  \includegraphics[height=4cm,width=0.48\textwidth]{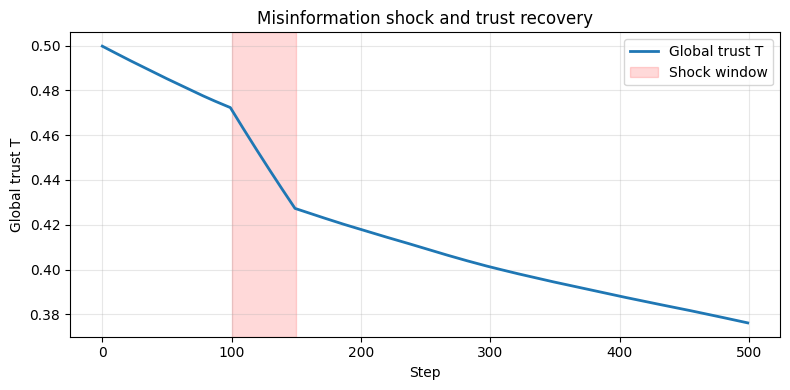}\hfill
  \includegraphics[height=4cm,width=0.48\textwidth]{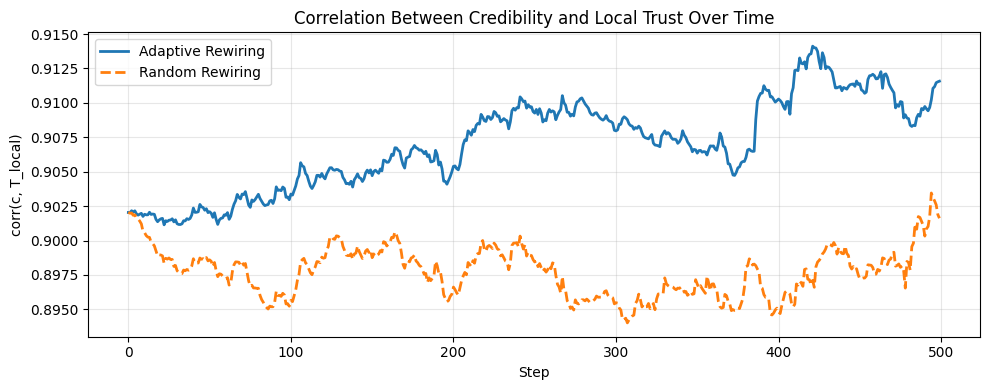}\\[3mm]
  % Row 2: Moran’s I | Assortativity
  \includegraphics[height=4cm,width=0.48\textwidth]{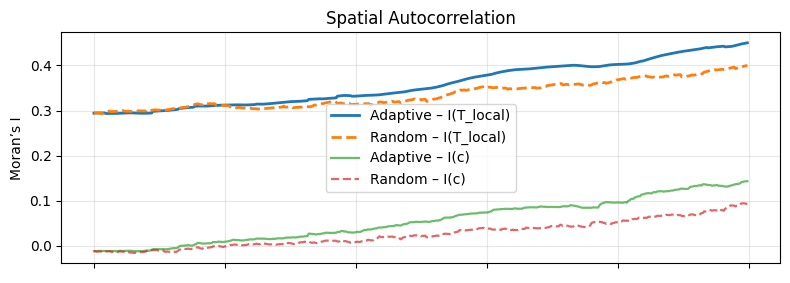}\hfill
  \includegraphics[height=4cm,width=0.48\textwidth]{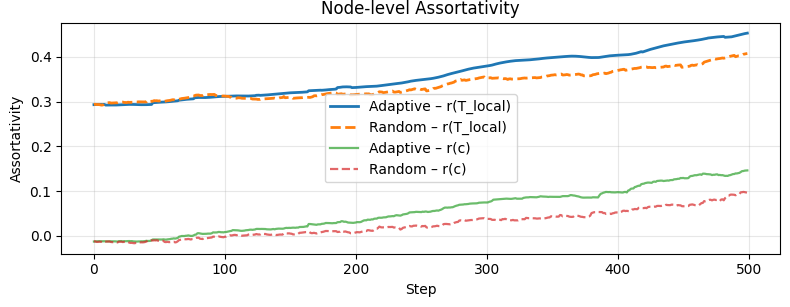}\\[3mm]
  % Row 3: Var(c) | Var(T_local)
  \includegraphics[height=4cm,width=0.48\textwidth]{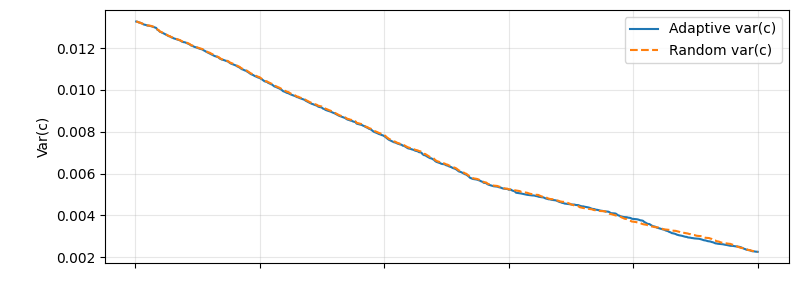}\hfill
  \includegraphics[height=4cm,width=0.48\textwidth]{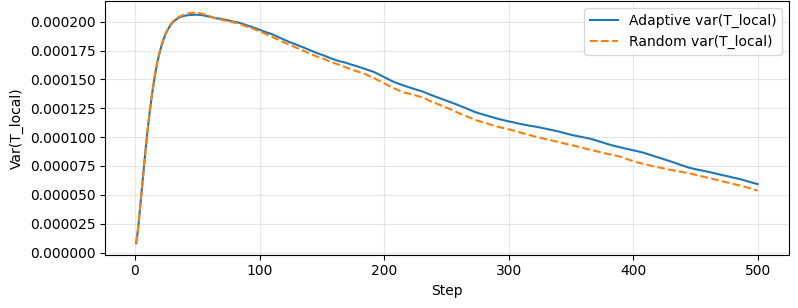}
  \caption{\textbf{Row 1 (shock test):} Global trust trajectory with shock window shaded (left); credibility--local-trust correlation (right). \textbf{Row 2 (spatial structure):} Spatial autocorrelation Moran’s $I$ (left); node-level assortativity (right) - adaptive rewiring (solid) vs.\ random rewiring (dashed). \textbf{Row 3 (variance):} $\mathrm{Var}(c)$ (left); $\mathrm{Var}(T_i)$ (right) - adaptive vs.\ random rewiring.}
  \label{fig:misinfo_shock}%
  \label{fig:network_stress}
\end{figure*}

\paragraph{Misinformation shock test.}
We inject a temporary surge of misinformation exposure and examine whether trust declines, attention reallocates toward misinformation, and the
system moves toward a lower-trust region (Eq.~\ref{eq:trust} coupled to Eq.~\ref{eq:prices}). Figure~\ref{fig:misinfo_shock} shows the intended
cascade: trust drops during/after the shock window, attention shifts toward misinformation, and the $T$--$\bar{c}$ trajectory moves
into a lower-trust/lower-credibility basin, consistent with a lock-in effect induced by trust-dependent costs.

\paragraph{Network stress test: adaptive vs.\ random rewiring.}
In the model, agents occasionally drop a connection to a dissimilar neighbor and form a new one with someone more like themselves, a process called \emph{adaptive rewiring}.
To check whether this actually matters (rather than being just routine network churn), we run the same scenario twice: once with adaptive rewiring (agents prefer similar neighbors) and once with random rewiring (new connections are chosen at random).
If the two runs look the same, the similarity-seeking behavior adds nothing.
Figure~\ref{fig:network_stress} shows they do not look the same: adaptive rewiring produces tighter opinion clusters (higher Moran’s $I$, higher assortativity) and a stronger link between an agent’s own credibility beliefs and those of their neighbors.
In plain terms, when agents actively seek out like-minded connections, the network sorts itself into echo chambers that reinforce and preserve local differences, an effect that random reshuffling cannot replicate.

\paragraph{Sensitivity / parameter dependence.}
Finally, we check whether different outcome families respond to different parts of the coupled system. Figs.~\ref{fig:validation_panel_3x5}--\ref{fig:validation_panel_r3} (Appendix~\ref{sec:appendix_sensitivity}) show
that (i) global trust level responds primarily to repair--harm balance (Eq.~\ref{eq:trust}), (ii) local trust dispersion is driven by heterogeneity
and network sorting (Eq.~\ref{eq:localtrust} plus rewiring/homophily), and (iii) preference dispersion reflects the interplay of imitation, noise,
and adaptive connectivity (Eq.~\ref{eq:eta} plus rewiring).

\subsection{Phase structure and regime separation}
\begin{figure*}[!t]
  \centering
  \includegraphics[width=0.48\textwidth]{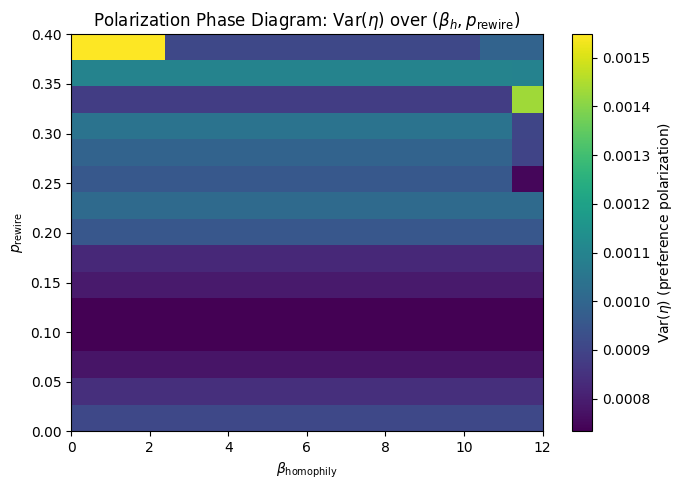}\hfill
  \includegraphics[width=0.48\textwidth]{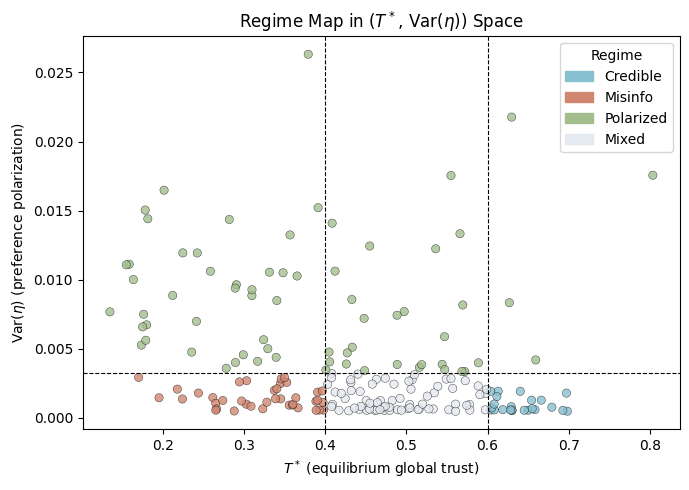}
  \caption{Left: polarization phase diagram $\mathrm{Var}(\eta)$ over $(\beta_{\mathrm{homophily}},p_{\mathrm{rewire}})$. Right: regime map in $T^*$--$\mathrm{Var}(\eta)$ space with threshold lines used to label outcomes.}

  \label{fig:phase_and_regime}
\end{figure*}

We characterize where polarization appears in parameter space and how regimes separate in outcome space.
Figure~\ref{fig:phase_and_regime} (left) maps preferences as variance Var(or $\eta$) over $(\beta_{\mathrm{homophily}},p_{\mathrm{rewire}})$, showing horizontal banding across
$p_{\mathrm{rewire}}$ and a peak at intermediate rewiring: too little rewiring fails to sort ties, while high-churn rewiring disrupts persistent
communities, limiting the time available for imitation to lock in divergence. Figure~\ref{fig:phase_and_regime} (right) shows that regimes separate along
two axes: equilibrium trust $T^*$ distinguishes credible versus misinformation outcomes, while $\mathrm{Var}(\eta)$ distinguishes integrated versus polarized
outcomes. Notably, polarized runs occur across a range of $T^*$, indicating that fragmentation can arise without global trust collapse when homophily and
rewiring reduce cross-cutting exposure.

\section{Assumptions and Limitations}
\label{sec:assumptions_limitations}

This paper is a mechanism-focused simulation study: the model is intentionally minimal so that the feedback structure is explicit and experimental comparisons admit clear attribution. We abstract the information environment into two competing streams (credible vs.\ misinformation) and represent trust as a single global stock $T(t)\in[0,1]$ updated by aggregate exposure (Eq.~\ref{eq:trust}). This implements an Ostrom-style depletion/replenishment logic, but it does not distinguish trust in institutions, media, peers, or domain-specific expertise, nor does it capture mixed-quality content, multiple topics, or strategic production of narratives. Attention allocation is modeled via trust-conditioned cognitive prices (Eq.~\ref{eq:prices}) under a unit budget; this operationalizes an attention-economy channel in which declining trust increases the effective cost of processing credible information, but it omits other determinants of attention such as affect, novelty, platform ranking, and adversarial manipulation. Credibility learning follows bounded confidence (Eqs.~\ref{eq:bc}--\ref{eq:cred_update}) and preferences evolve via imitation with noise (Eq.~\ref{eq:eta}); these are standard stylizations, but they do not capture motivated reasoning, memory, identity-protective cognition, or strategic behavior.

Network structure is represented with a small-world substrate and adaptive rewiring based on preference similarity. This captures co-evolution of exposure pathways and beliefs, but it abstracts away platform-mediated exposure mechanisms (recommendation systems, moderation) and richer forms of network change such as group entry/exit, multiplex ties, and directed influence. As a result, the regimes reported here should be interpreted as qualitative operating modes of a simplified coupled system rather than as forecasts for a particular platform.

Finally, the evaluation scope emphasizes regime characterization, baseline reductions, perturbation and stress tests, and parameter dependence diagnostics (Section~\ref{sec:results}). We do not calibrate parameters to observational data, and we treat validation as qualitative pattern checking rather than predictive accuracy. Regime boundaries can be sensitive to initialization and stochasticity; therefore, phase maps should be interpreted as empirical summaries over explored parameter ranges, best supported by replicated, uncertainty-aware comparisons. Within these limits, the model’s intended use is comparative and experimental: to isolate which feedback channels govern which macroscopic outcomes

% ============================================================
\section{Conclusions}
\label{sec:conclusion}

This paper lays a simulation foundation for studying how misinformation propagates in an \emph{attention economy} while keeping an Ostrom-style commons analysis at the center of the modeling frame. The core claim is not simply that falsehoods spread, but that attention allocation creates a population-level externality: when aggregate attention shifts toward low-credibility content, it depletes the shared conditions under which claims can be evaluated. We operationalize that commons logic with a global trust stock that is repaired or harmed by aggregate exposure (Eq.~\ref{eq:trust}), and we link it to individual behavior through trust-conditioned cognitive prices that shape attention allocation (Eq.~\ref{eq:prices}) and through local social learning on an adaptive network (Fig.~\ref{fig:model_overview}). In this way, the model treats trust as a shared enabling resource (commons) and attention as a privately allocated budget whose distribution governs the commons’ replenishment or depletion.

Across simulation experiments, this coupling produces multiple qualitative operating modes—credible stability, misinformation dominance, polarization, and a mixed baseline regime—with consistent signatures in trajectories, distributions, and network structure (Figs.~\ref{fig:panel_networks} and \ref{fig:panel_metrics}; Table~\ref{tab:regime_params}). The regimes clarify how commons dynamics interact with attention incentives: (i) when repair dominates harm, the trust stock remains high and the cost structure favors credible attention, supporting stable credibility dynamics; (ii) when harm dominates, declining trust makes credible processing increasingly costly, shifting attention toward misinformation and creating a self-reinforcing degradation pathway; and (iii) under intermediate repair/harm, adaptive network structure can transform moderate heterogeneity into persistent clustered environments, generating polarization even without assuming intrinsically polarized agents.

A central implication for commons governance is that \emph{trust level} and \emph{fragmentation} are not the same control problem. Repair--harm balance primarily governs whether the system converges to a high-trust equilibrium or collapses (Fig.~\ref{fig:panel_metrics}, Row~1), whereas homophily and rewiring govern whether exposure remains cross-cutting or segregates into reinforcing clusters (Fig.~\ref{fig:panel_networks}). The phase and regime maps show that polarization can arise across a range of equilibrium trust levels when cross-cutting exposure erodes (Fig.~\ref{fig:phase_and_regime}), suggesting that interventions that merely ``raise trust'' may still fail if the exposure structure continues to sort attention into separate epistemic neighborhoods. Put differently, an Ostrom-style framing motivates two distinct families of levers: those that shift the replenishment/depletion balance of the shared stock (repair, harm reduction), and those that preserve or restore cross-cutting conditions for monitoring, sanctioning, and shared evidence (structural exposure, network mixing).

The computational contribution is a modular testbed for designed experiments on these governance levers. Section~\ref{sec:results} reports standard simulation-study diagnostics that support interpretability and reproducibility, including baseline reductions, implementation invariants, perturbation experiments, stress tests, and parameter-dependence analyses (e.g., Figs.~\ref{fig:verification_baselines_3x2}--\ref{fig:validation_panel_3x5}). These diagnostics help attribute regime behavior to specific feedback channels rather than to incidental implementation choices.

Future work will extend this foundation in three directions. First, we will replace the binary credible/misinformation split with richer, multi-topic and mixed-quality streams, enabling more realistic attention competition. Second, we will ground trust signals and network substrates in empirical data (e.g., platform networks, community structure, measured credibility cues) to support calibration and out-of-sample checks. Third, we will use the testbed to run intervention experiments aligned with commons design principles—separately targeting stock dynamics (repair/harm) and exposure structure (rewiring/homophily)—to evaluate which governance strategies stabilize the epistemic commons under attention scarcity.

\section*{Acknowledgments}
I thank my advisor, Dr. Robert Axtell, and colleagues for helpful discussions and feedback on the model design and experiments. I especially want to thank Dr. Hamdi Kavak for his support as a member of my dissertation committee and his class (Verification and Validation of Models), where this project took most of its shape. I also thank the anonymous reviewers for their comments, which improved clarity and framing.

This manuscript benefited from the use of generative AI tools for editorial assistance (e.g., wording, rephrasing, and LaTeX formatting suggestions). All technical content, modeling choices, experiments, and interpretations were developed and verified by the author, who takes full responsibility for the paper. No proprietary or sensitive information was provided to the AI tools, and the tools were not used to generate or run simulations or to produce experimental results.

\newpage

% Please don't change the bibliographystyle style
\bibliographystyle{scsproc}

% AUTHOR: Include your bib file here
\bibliography{assets/demobib}

\section*{Author Biographies}

\textbf{\uppercase{Vrinda Malhotra}} is a PhD student in the Computational Data Science Department at George Mason University, Virginia, USA. Her main research includes using computational methods to study the spread of misinformation and to build systemic resilience against it. Her email is \email{vmalhot2@gmu.edu}.

\newpage

\appendix

\section{Full Network Panel (Panel Set A)}
\label{sec:appendix_networks}
% (Panel Set B, sensitivity panel, and unit tests follow below)

% Panel Set A — Row 1: Credibility
\begin{figure*}[p]
  \centering
  \includegraphics[width=0.48\textwidth]{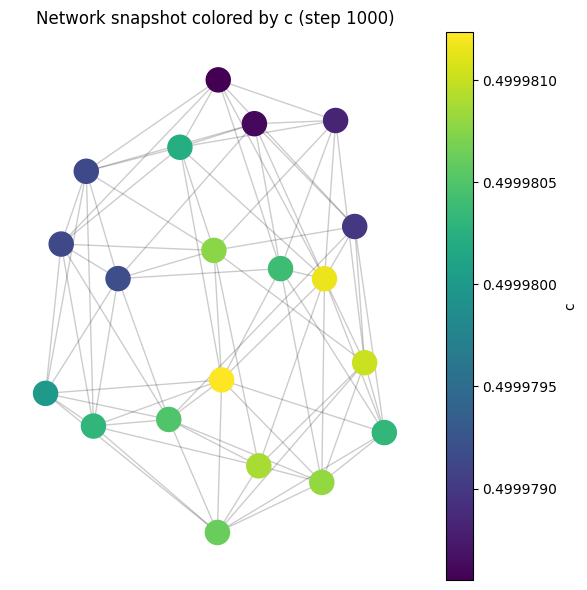}\hfill
  \includegraphics[width=0.48\textwidth]{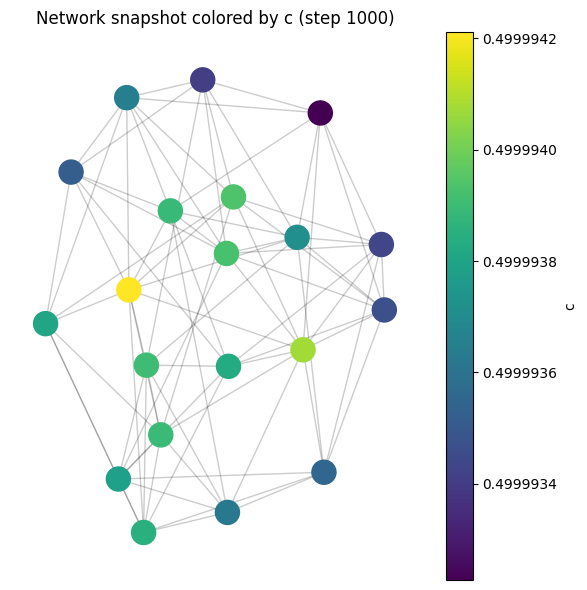}\\[4mm]
  \includegraphics[width=0.48\textwidth]{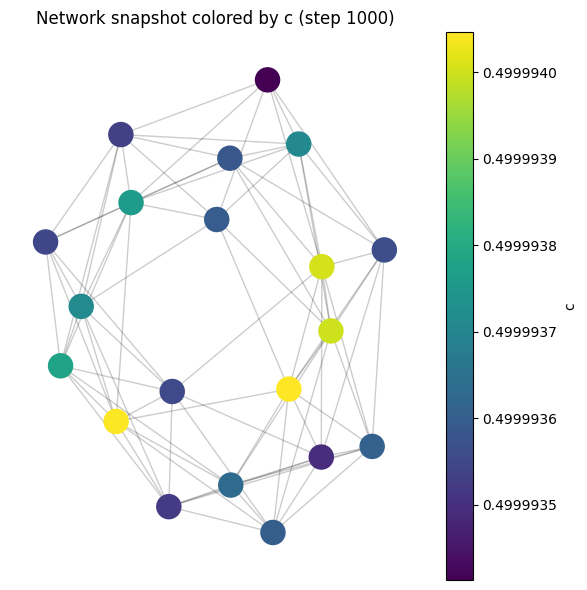}\hfill
  \includegraphics[width=0.48\textwidth]{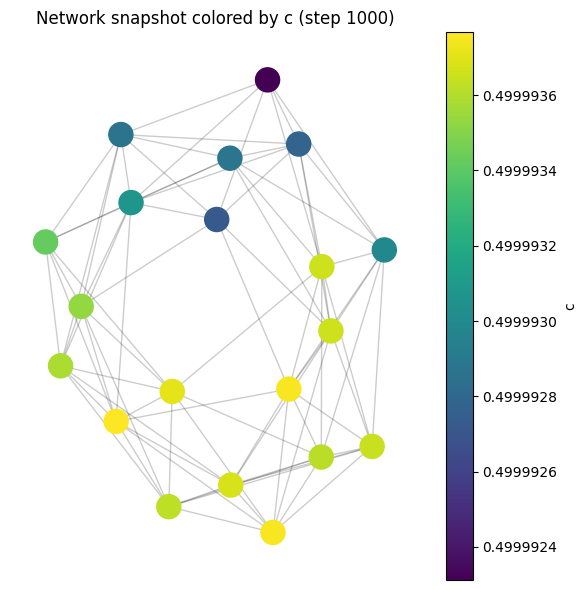}
  \caption{Panel Set A, Row~1 (credibility $c_i$): network snapshots for (A) Credible (top-left), (B) Misinformation (top-right), (C) Polarized (bottom-left), (D) Baseline (bottom-right). Reproduced at smaller scale in the main body (Fig.~\ref{fig:panel_networks}).}
  \label{fig:panel_networks_full}
\end{figure*}

% Panel Set A — Row 2: Local trust
\begin{figure*}[p]
  \centering
  \includegraphics[width=0.48\textwidth]{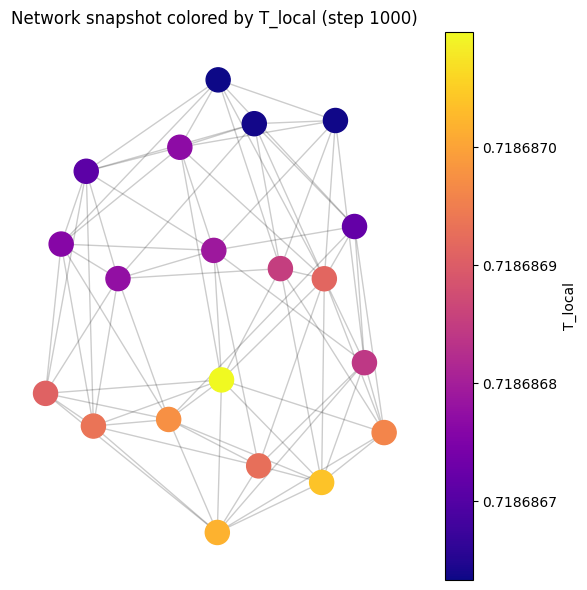}\hfill
  \includegraphics[width=0.48\textwidth]{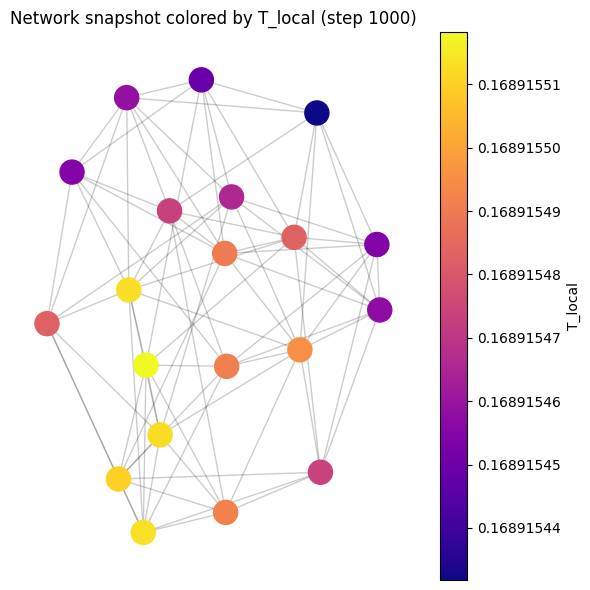}\\[4mm]
  \includegraphics[width=0.48\textwidth]{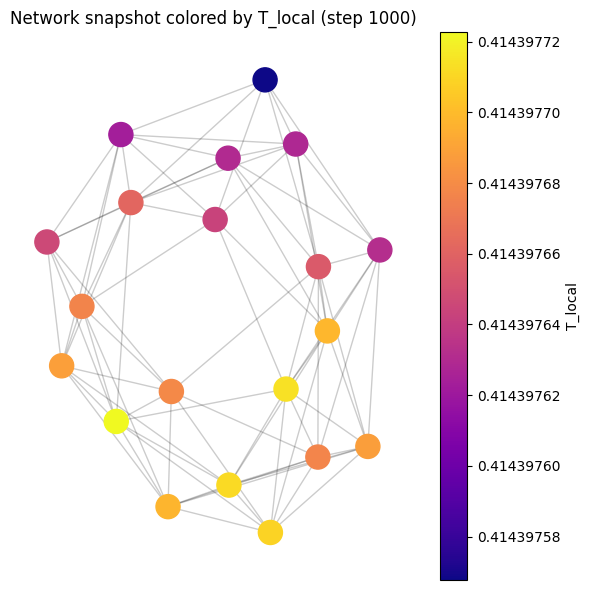}\hfill
  \includegraphics[width=0.48\textwidth]{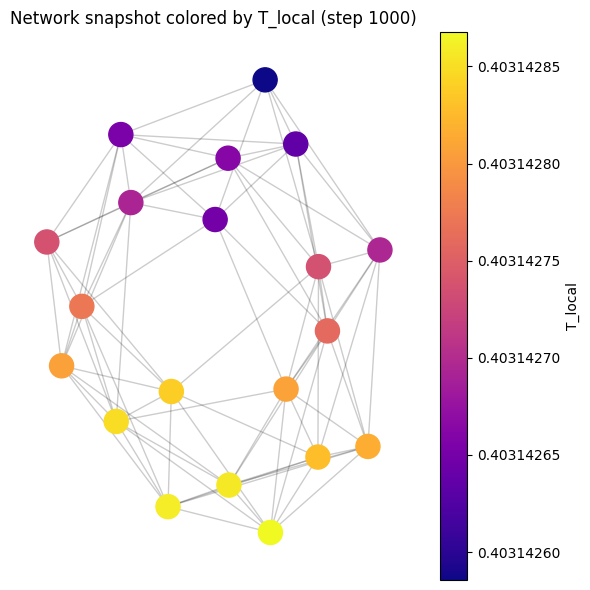}
  \caption{Panel Set A, Row~2 (local trust $T_i$): network snapshots for (A) Credible (top-left), (B) Misinformation (top-right), (C) Polarized (bottom-left), (D) Baseline (bottom-right).}
  \label{fig:panel_networks_full_r2}
\end{figure*}

% Panel Set A — Row 3: Preference
\begin{figure*}[p]
  \centering
  \includegraphics[width=0.48\textwidth]{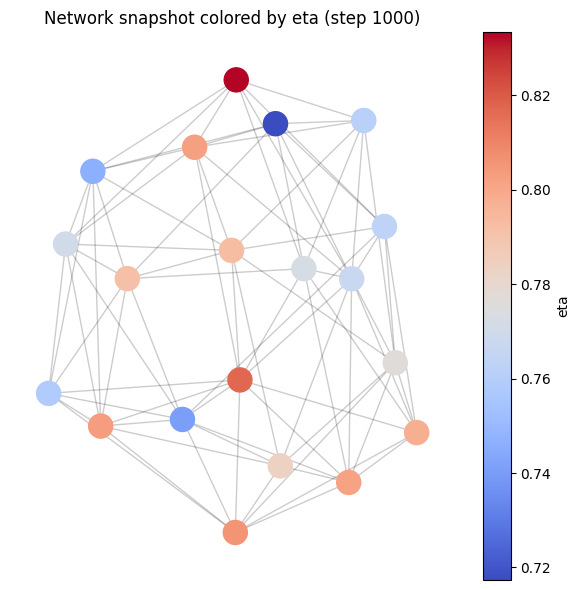}\hfill
  \includegraphics[width=0.48\textwidth]{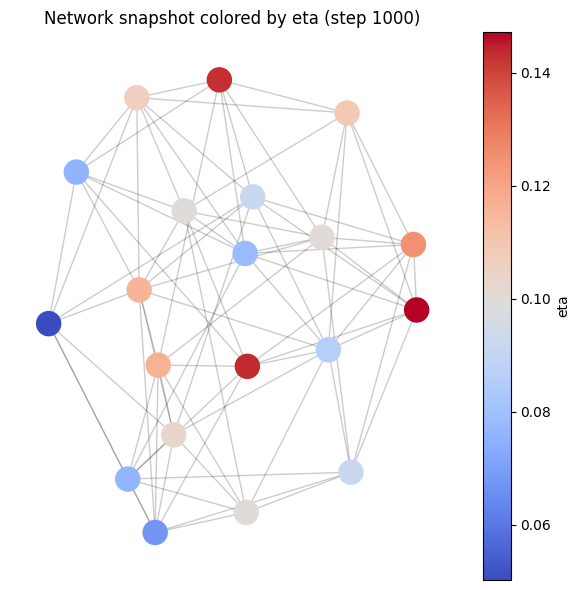}\\[4mm]
  \includegraphics[width=0.48\textwidth]{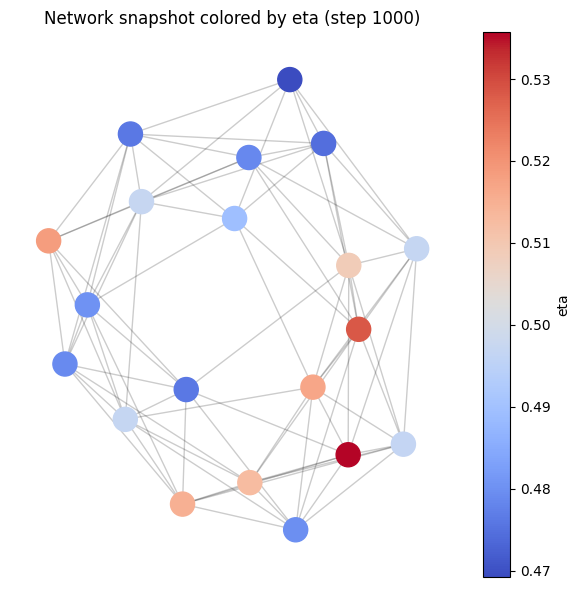}\hfill
  \includegraphics[width=0.48\textwidth]{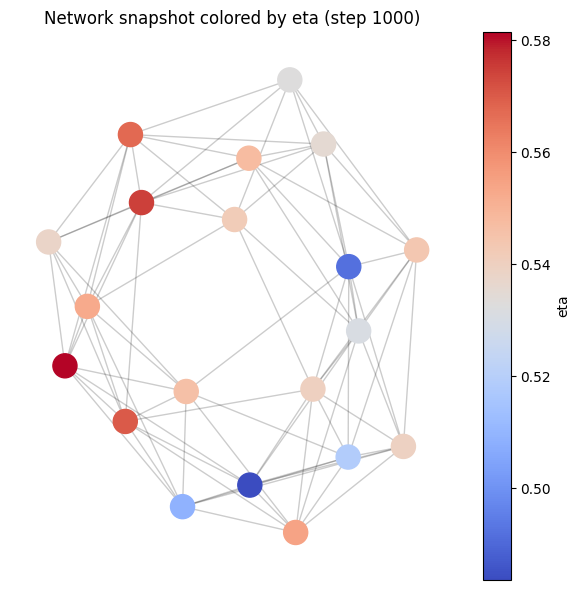}
  \caption{Panel Set A, Row~3 (preference $\eta_i$): network snapshots for (A) Credible (top-left), (B) Misinformation (top-right), (C) Polarized (bottom-left), (D) Baseline (bottom-right).}
  \label{fig:panel_networks_full_r3}
\end{figure*}

% =========================
% Appendix: Panel Set B (rows 1-3)
% =========================
\section{Full Non-Network Panel (Panel Set B, Rows 1--3)}
\label{sec:appendix_panel_b}

% Panel Set B — Row 1: Trust and attention trajectories
\begin{figure*}[p]
  \centering
  \includegraphics[width=0.48\textwidth,height=9cm,keepaspectratio]{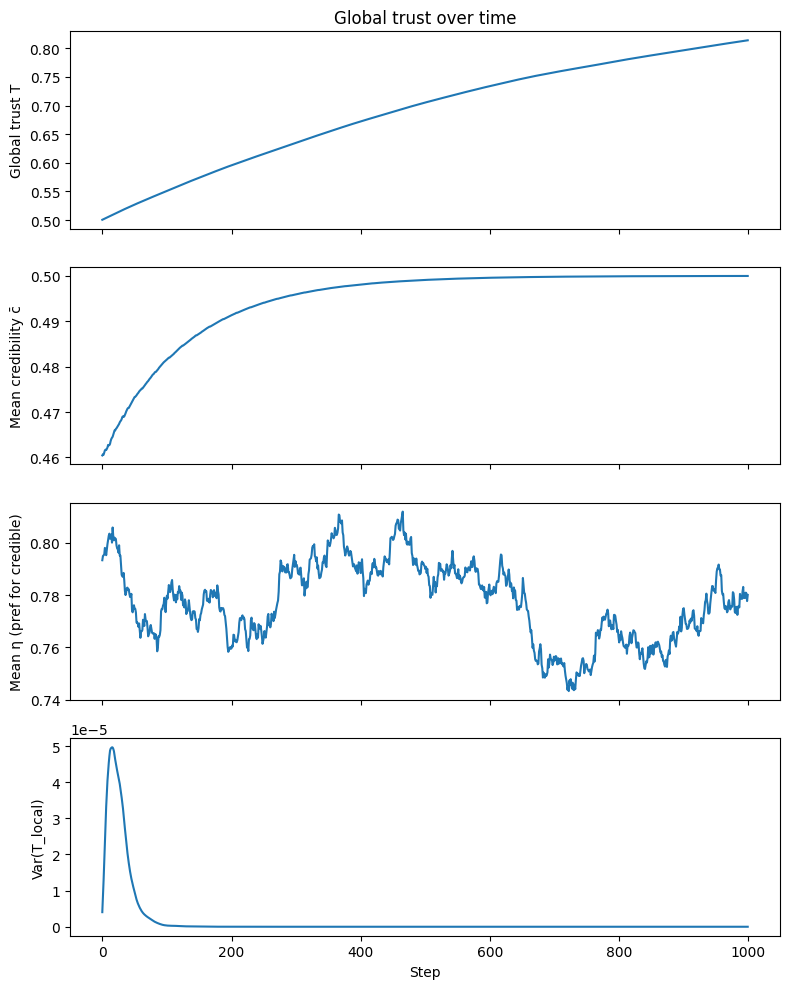}\hfill
  \includegraphics[width=0.48\textwidth,height=9cm,keepaspectratio]{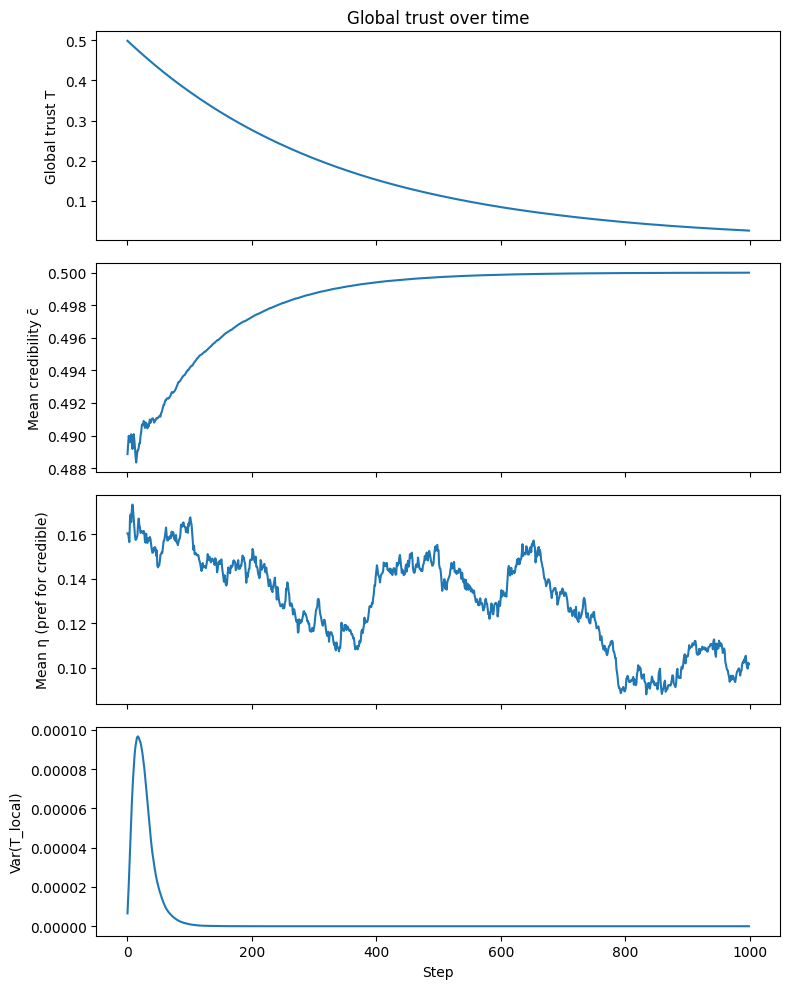}\\[4mm]
  \includegraphics[width=0.48\textwidth,height=9cm,keepaspectratio]{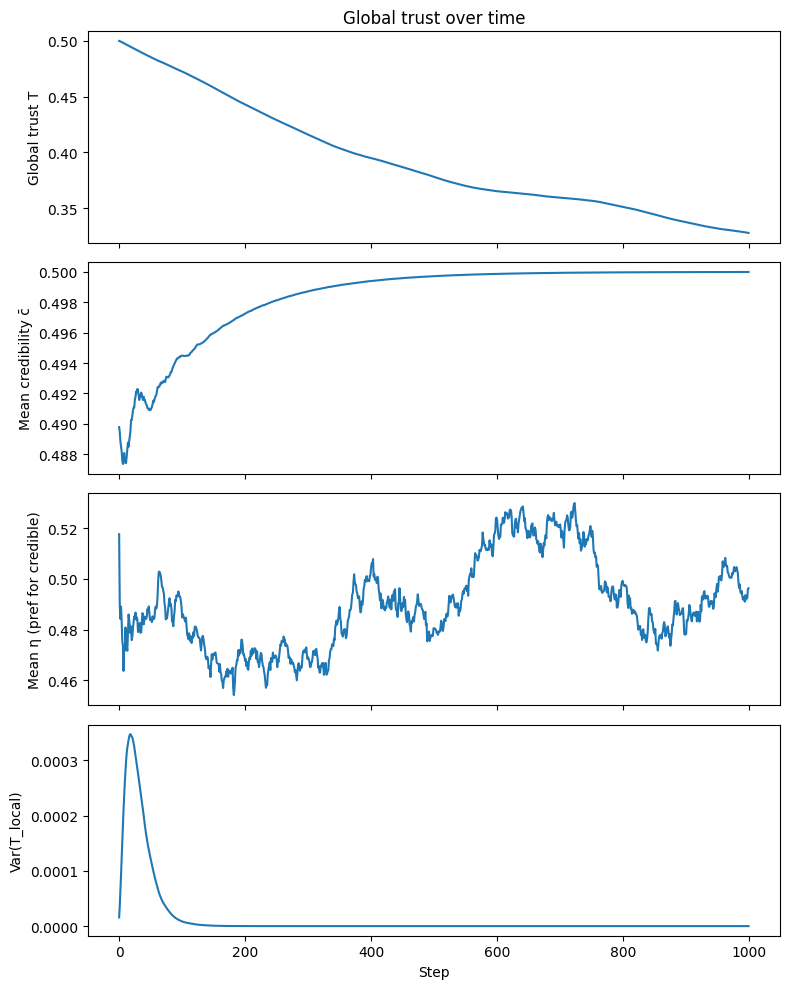}\hfill
  \includegraphics[width=0.48\textwidth,height=9cm,keepaspectratio]{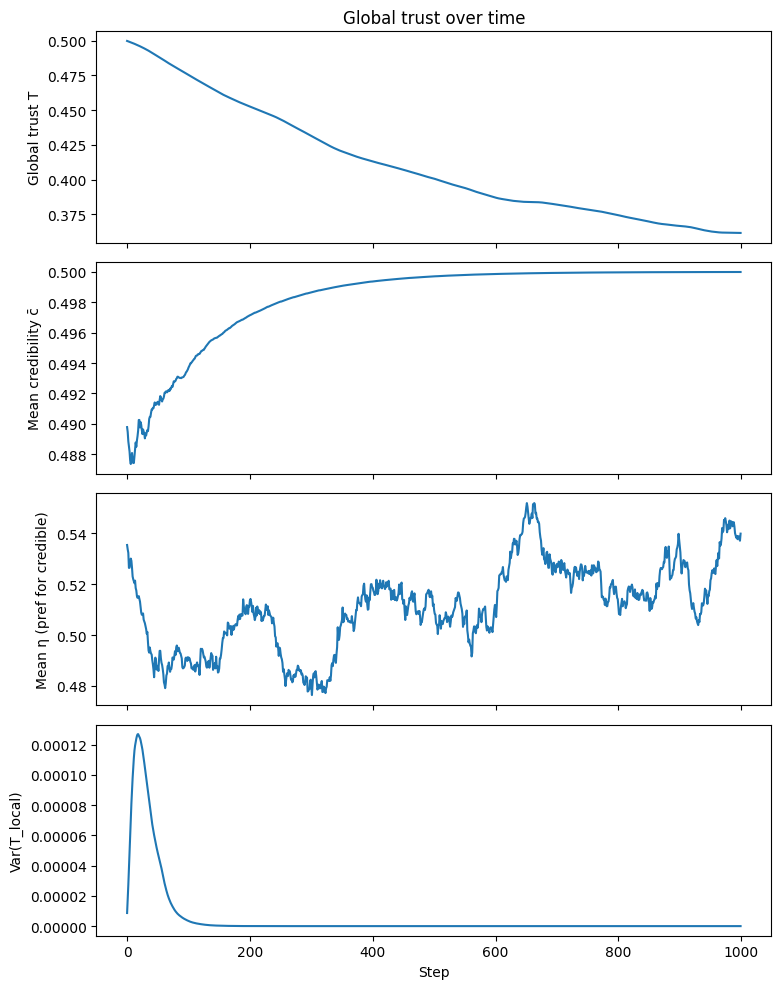}
  \caption{Panel Set B, Row~1 (trust and attention trajectories): (A) Credible (top-left), (B) Misinformation (top-right), (C) Polarized (bottom-left), (D) Baseline (bottom-right). Row~4 (validation diagnostics) is in the main body (Fig.~\ref{fig:panel_metrics}).}
  \label{fig:panel_metrics_full}
\end{figure*}

% Panel Set B — Row 2: Distributions
\begin{figure*}[p]
  \centering
  \includegraphics[width=0.48\textwidth]{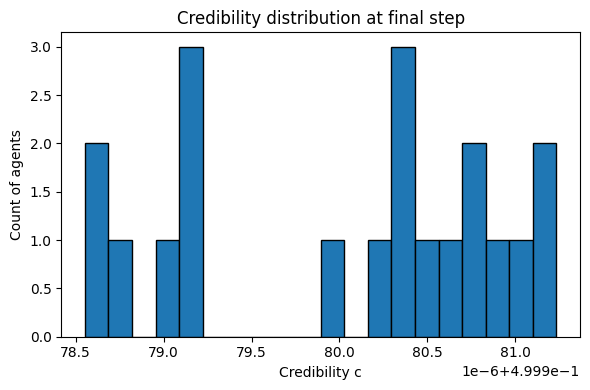}\hfill
  \includegraphics[width=0.48\textwidth]{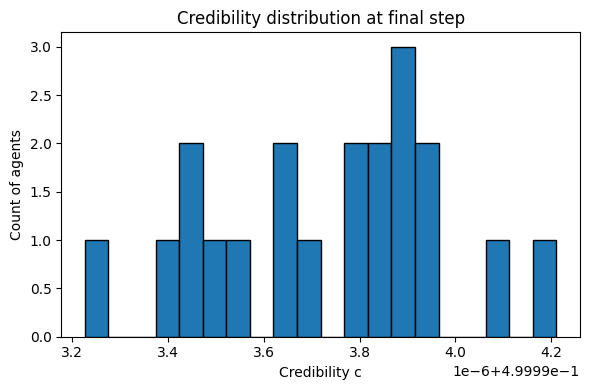}\\[4mm]
  \includegraphics[width=0.48\textwidth]{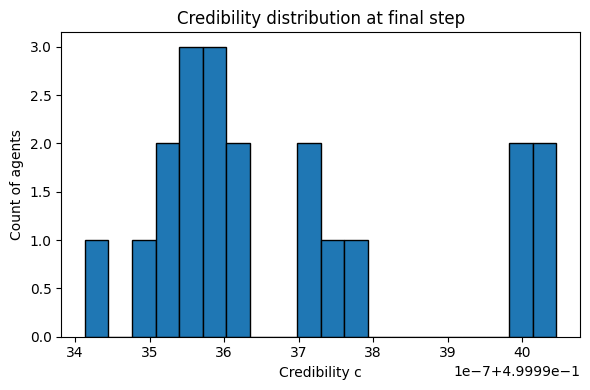}\hfill
  \includegraphics[width=0.48\textwidth]{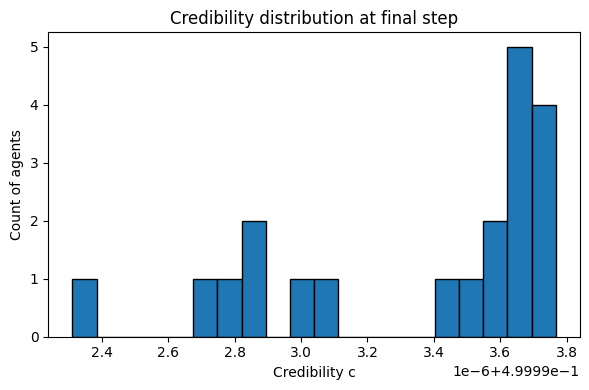}
  \caption{Panel Set B, Row~2 (credibility and preference distributions): (A) Credible (top-left), (B) Misinformation (top-right), (C) Polarized (bottom-left), (D) Baseline (bottom-right).}
  \label{fig:panel_metrics_full_r2}
\end{figure*}

% Panel Set B — Row 3: Verification baselines
\begin{figure*}[p]
  \centering
  \includegraphics[width=0.48\textwidth]{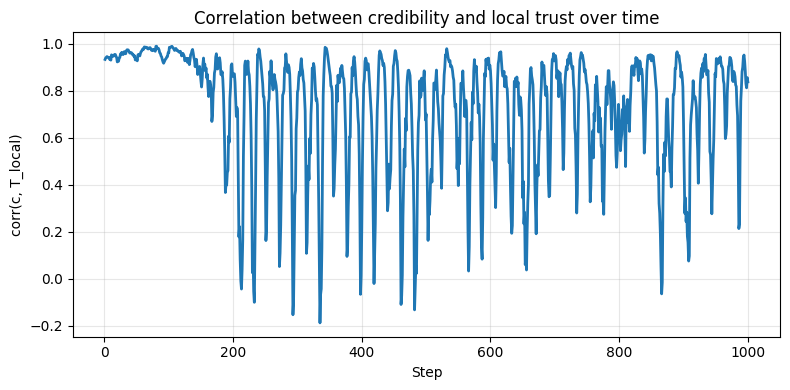}\hfill
  \includegraphics[width=0.48\textwidth]{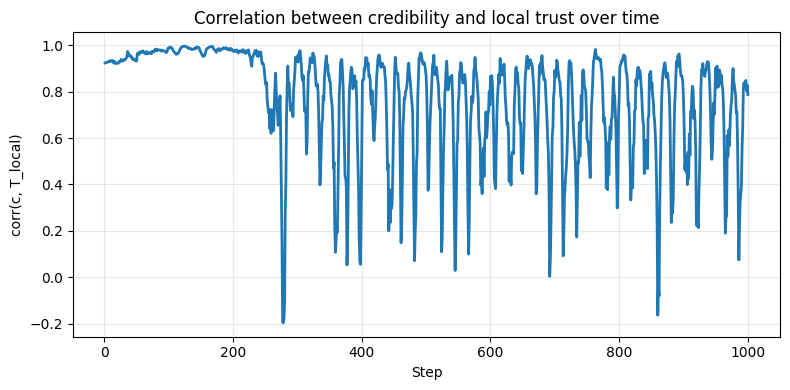}\\[4mm]
  \includegraphics[width=0.48\textwidth]{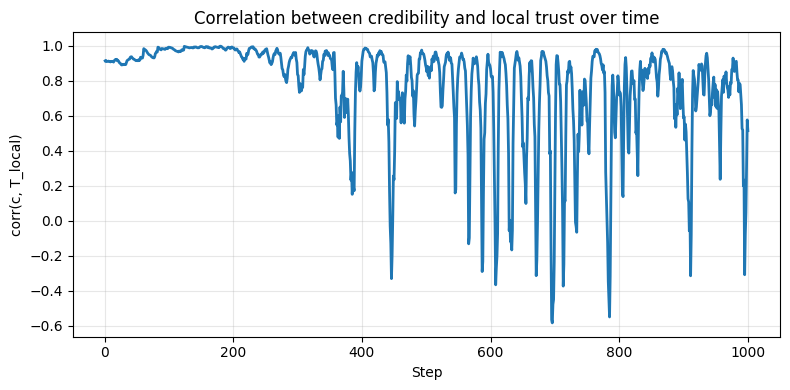}\hfill
  \includegraphics[width=0.48\textwidth]{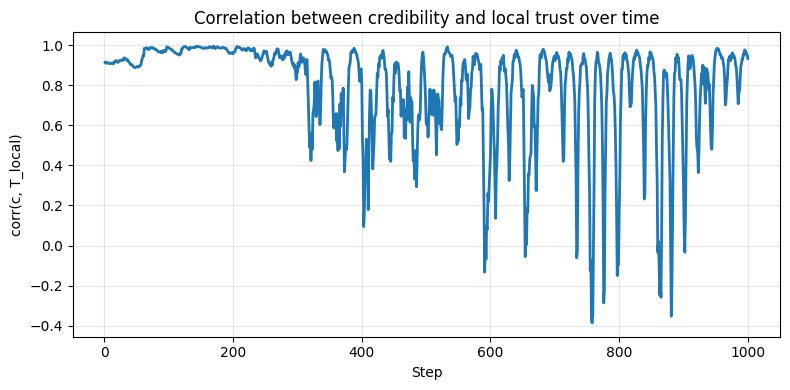}
  \caption{Panel Set B, Row~3 (verification baselines): (A) Credible (top-left), (B) Misinformation (top-right), (C) Polarized (bottom-left), (D) Baseline (bottom-right).}
  \label{fig:panel_metrics_full_r3}
\end{figure*}

% Panel Set B — Row 4: Global trust and mean credibility line plots
\begin{figure*}[p]
  \centering
  \includegraphics[width=0.48\textwidth]{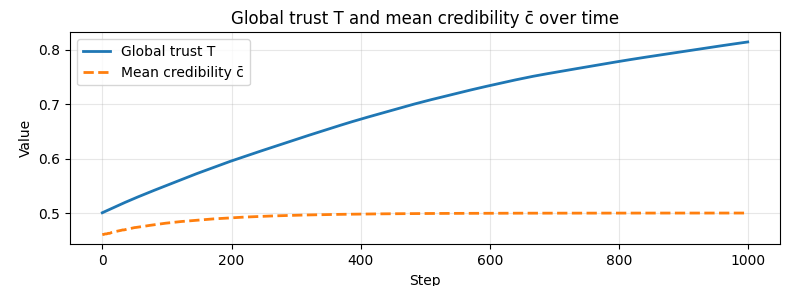}\hfill
  \includegraphics[width=0.48\textwidth]{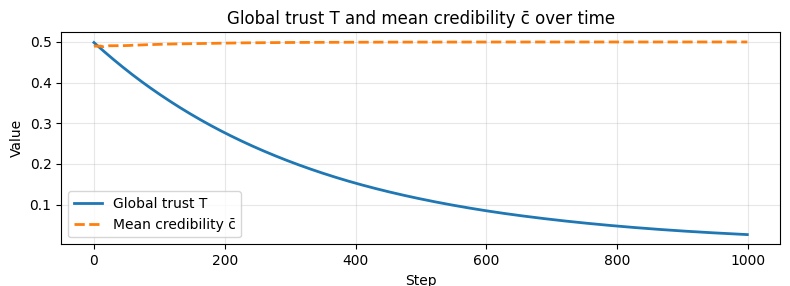}\\[4mm]
  \includegraphics[width=0.48\textwidth]{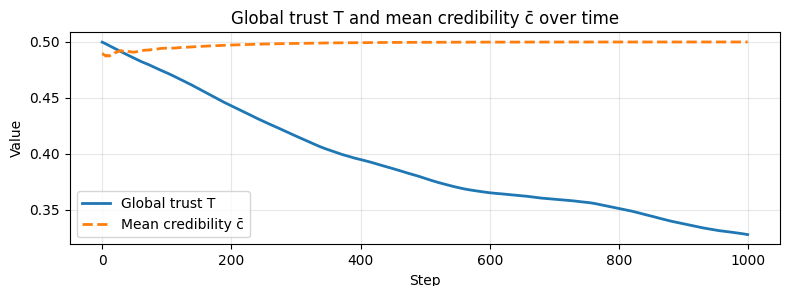}\hfill
  \includegraphics[width=0.48\textwidth]{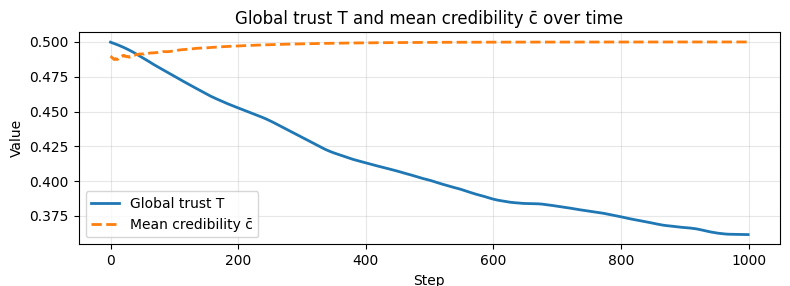}
  \caption{Panel Set B, Row~4 (global trust $T$ and mean credibility $\bar{c}$ over time): (A) Credible (top-left), (B) Misinformation (top-right), (C) Polarized (bottom-left), (D) Baseline (bottom-right). Phase plots for the same runs appear in the main body (Fig.~\ref{fig:panel_metrics}).}
  \label{fig:panel_phase_full}
\end{figure*}

% =========================
% Appendix: Sensitivity panel
% =========================
\section{Parameter Sensitivity Panel}
\label{sec:appendix_sensitivity}

% Sensitivity — Row 1: Global trust T*
\begin{figure*}[p]
  \centering
  \includegraphics[width=\textwidth]{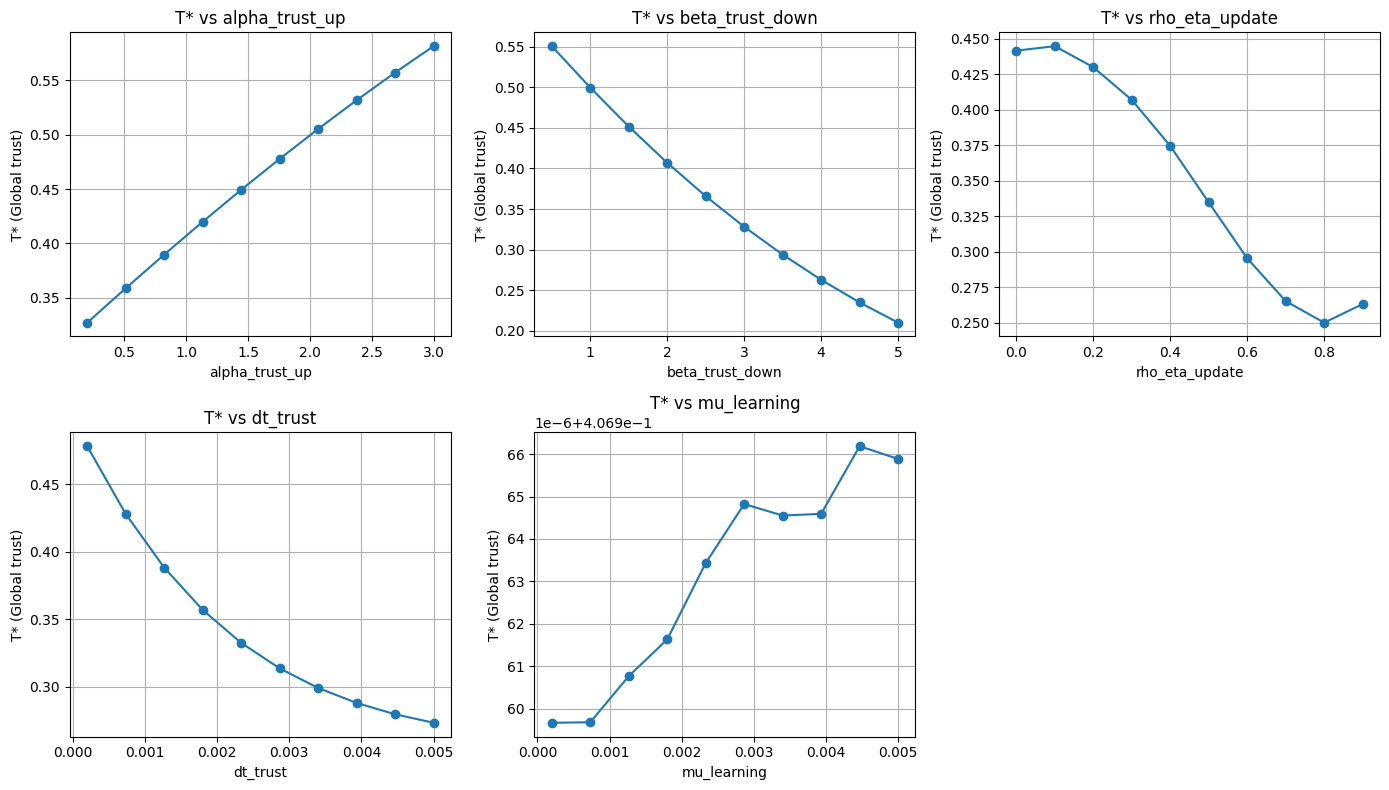}
  \caption{Parameter sensitivity, Row~1: global trust level $T^*$ as a function of the parameters with the strongest observed influence (one parameter varied per panel).}
  \label{fig:validation_panel_3x5}
\end{figure*}

% Sensitivity — Row 2: Local trust dispersion
\begin{figure*}[p]
  \centering
  \includegraphics[width=\textwidth]{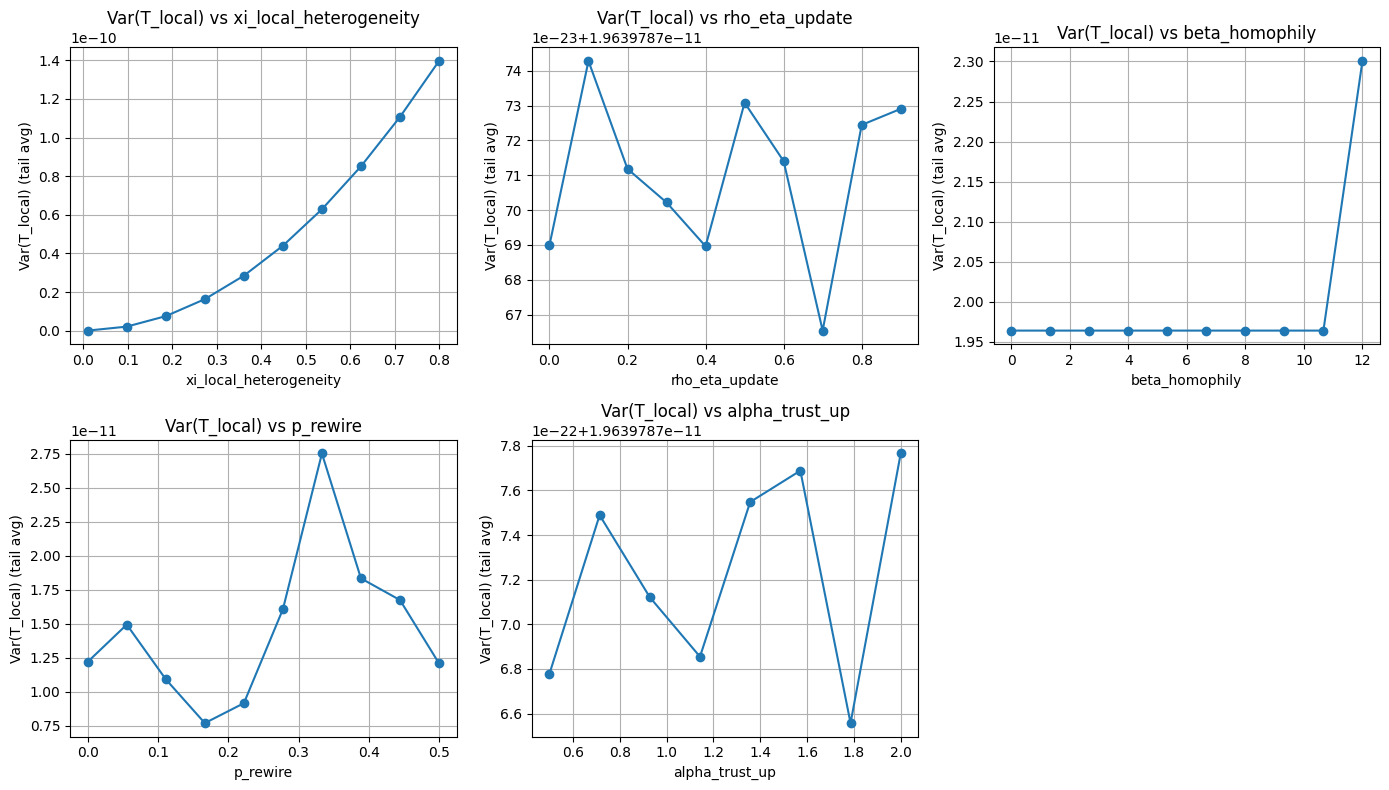}
  \caption{Parameter sensitivity, Row~2: local trust dispersion $\mathrm{Var}(T_i)$ as a function of the parameters with the strongest observed influence (one parameter varied per panel).}
  \label{fig:validation_panel_r2}
\end{figure*}

% Sensitivity — Row 3: Preference dispersion
\begin{figure*}[p]
  \centering
  \includegraphics[width=\textwidth]{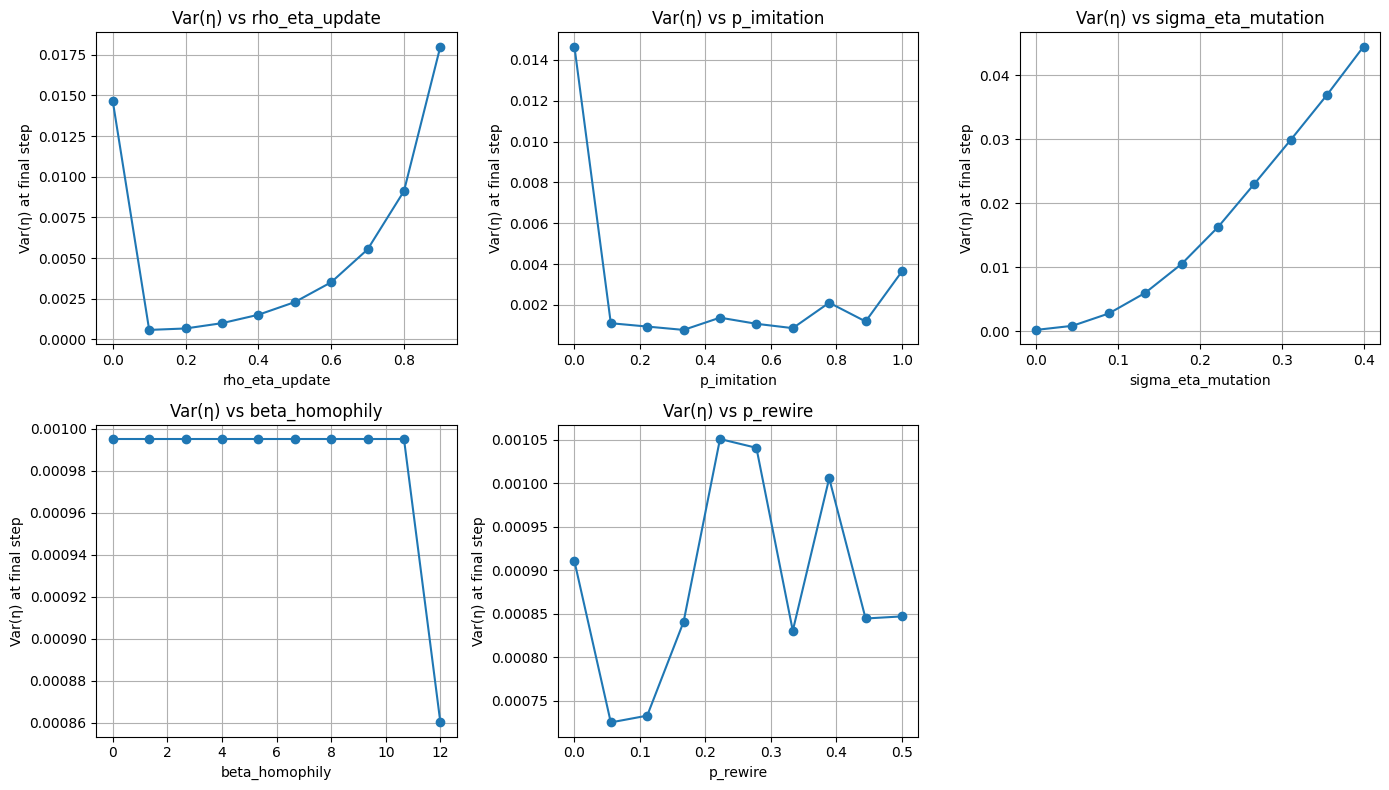}
  \caption{Parameter sensitivity, Row~3: preference dispersion $\mathrm{Var}(\eta_i)$ as a function of the parameters with the strongest observed influence (one parameter varied per panel).}
  \label{fig:validation_panel_r3}
\end{figure*}

% =========================
% Appendix: Unit tests
% =========================
\section{Verification Unit Tests}
\label{sec:appendix_tests}

\begin{table}[h]
  \caption{Verification unit tests (pytest) grouped by model component.}
  \label{tab:verification_tests}
  \centering
  \small
  \setlength{\tabcolsep}{5pt}
  \renewcommand{\arraystretch}{1.25}
  \begin{tabular}{p{0.19\linewidth}p{0.43\linewidth}p{0.31\linewidth}}
    \toprule
    Category & What it verifies & Representative test \\
    \midrule
    Credibility dynamics &
      Deffuant updates move beliefs toward each other within tolerance $\varepsilon$, leave them unchanged outside it, and keep $c_i\in[0,1]$. &
      \texttt{test\_bounded\_confidence\_moves\_toward\_each\_other} \\[2pt]
    Attention allocation &
      Attention shares are non-negative and saturate correctly at $\eta=0$ (all misinfo) and $\eta=1$ (all credible). &
      \texttt{test\_attention\_allocation\_respects\_preference\_extremes} \\[2pt]
    Price function &
      Credible-content price decreases monotonically with $T$; misinfo price increases; data collector returns correct shapes. &
      \texttt{test\_price\_monotonicity} \\[2pt]
    Global trust &
      Trust increases under pure credible exposure and decreases under pure misinfo, validating Eq.~(trust dynamics). &
      \texttt{test\_global\_trust\_increases\_when\_only\_credible\_attention} \\[2pt]
    Local trust &
      $\mathrm{Var}(T_i)>0$ when $\xi>0$; collapses to $T(t)$ when $\xi=0$ (homogeneous baseline). &
      \texttt{test\_local\_trust\_equals\_global\_when\_xi\_zero} \\[2pt]
    Network / rewiring &
      Edge count is preserved after rewiring; edges change when rewiring probability is high. &
      \texttt{test\_rewiring\_preserves\_number\_of\_edges} \\
    \bottomrule
  \end{tabular}
\end{table}

\end{document}